# Ambient and Pressure Dependent Superconductivity with Hydrogen Storage Potential in Quaternary Hydride $LiMgZr_2H_{12}$: A Comprehensive First-principles Insights

**Jubair Hossan Abir[1], Tauhidur Rahman[1], Salauddin Muhammad Anis[1], Saleh Hasan Naqib[1*], Raihana Shams Islam[1*]**

[1]Department of Physics, University of Rajshahi, Rajshahi 6205
*Corresponding authors; Emails: salehnaqib@yahoo.com, rsislam@ru.ac.bd

## Abstract

Molecular hydrides have attracted relatively less attention in the search for high $T_c$ superconductors because their hydrogen quasi-molecular units tend to be electronically inactive for superconductivity. In contrast, hydrogen rich compounds under high pressure have been widely considered strong candidates for achieving room-temperature superconductivity. However, their dependence on extreme pressure conditions significantly constrains their practical applicability. This work investigates hydrogen-rich superconducting materials that may be stable under ambient pressure conditions. Motivated by recent studies on the $MgZrH_{2n}$ family, a $LiMgZr_2H_{12}$ structure with *Pmmm* symmetry was designed. The mechanical, thermodynamic, and dynamical stability of the compound, together with its electronic and optical properties, were systematically investigated using first-principles calculations. Li doping in $LiMgZr_2H_{12}$ significantly increases the hydrogen derived contribution near the Fermi level ($E_F$) and strengthens the electron-phonon coupling constant (λ) compared with $MgZrH_6$. $LiMgZr_2H_{12}$ exhibits a critical temperature of 72.76 K at ambient pressure, which is further enhanced by applying pressure. At 10 GPa the critical temperature increases to 77.3 K. Elastic property analysis shows that the material remains mechanically stable over the pressure range studied (0-10 GPa). It also behaves like a ductile material suitable for current carrying applications. The material has a high machinability index, which is much higher than that of stainless steels. In addition, $LiMgZr_2H_{12}$ exhibits a gravimetric hydrogen storage capacity of 5.36 wt%, indicating its potential as a promising candidate for hybrid hydrogen storage technologies. This work offers a new direction for designing high-$T_c$ hydrides at ambient conditions.

## 1 Introduction

One of the enduring challenges in condensed matter physics is the discovery of superconductors that can function at high temperatures. In superconductors, electricity can flow without any loss below a critical temperature, denoted by $T_c$. Superconductors have drawn considerable attention due to their wide range of possible technological applications. The search for materials that can exhibit superconductivity at high temperatures or even at room temperature is now a central goal for condensed matter physics and materials science [**1**]. Hydrogen-rich compounds under high pressures have been recognized as promising room-temperature superconductor candidates for a long time [**1**]. These types of compounds are binary polyhydrides with exceptionally high hydrogen content, such as *Im*3*m*-$H_3S$ ($T_c$ = 203 K) and $LaH_{10}$ ($T_c$ = 250-260 K) [**2-4**]. In general, the behavior of superhydrides can be described using the Migdal-Eliashberg theory of strong electron-phonon coupling [**2**]. Moreover, using the Ginzburg-Landau-Abrikosov-Gorkov (GLAG) theory, it has been seen that increasing the effective mass of electrons should increase the superconducting transition temperature even above room temperature [**3**]. More binary hydrides have been discovered, such as $LaH_{10}$ ($T_c$ = 250 K), $YH_6$ ($T_c$ = 220 K), $YH_9$ ($T_c$ = 243 K), and $CaH_6$ ($T_c$ = 215 K), which have superconducting critical temperatures greater than 200 K [**4–6**]. Despite these advances, most hydrogen-rich superconductors remain stable only at very high pressures. This makes their practical application limited. Accordingly, Qun Wei et al. have recently predicted a compound $LiMgZr_2H_{12}$ which may be a superconductor with a $T_c$ of 60.8 K at ambient pressure using Electron-phonon coupling (EPC) analysis [**1**]. This interesting prediction of achieving a stable high $T_c$ superconductor at ambient pressure motivated us to further study the $LiMgZr_2H_{12}$ compound under various pressures using first-principles calculations. In this work, we have explored the physical properties of $LiMgZr_2H_{12}$, which has an orthorhombic structure with *Pmmm* symmetry (space group No. 47). We have analyze the structural, elastic, electronic, bonding, thermal, acoustic, optical, phonon, and superconducting state properties of $LiMgZr_2H_{12}$ under pressure up to 10 GPa. To our knowledge, no pressure-dependent density functional theory-based work has been reported in the literature for this compound. It is important to note that many essential physical properties of $LiMgZr_2H_{12}$ are neither well known experimentally nor well known computationally in comparison to other binary and ternary hydrogen-rich compounds. Qun Wei et al. have taken the Mg-Zr-H system for modifying it by doping Li [**1**]. Their motivation for selecting this system was that in the $MgZrH_{2n}$ family, for $MgZrH_6$, a $T_c$ value of 80.3 K at 36 GPa has been predicted, which was an indication of significant potential for superconductivity in related compounds [**7**]. Qun Wei et al. have introduced Li to reduce the pressure necessary for structural stabilization. The atomic size of Li is similar to Mg, but its lower electronegativity makes it transfer more charge to H [**1**].

As far as we know, a significant number of the physical features, such as elastic, energy-dependent optical properties, thermal properties, acoustic properties, phonon dispersions, electronic energy density of states (DOS) at the Fermi level, strength of EPC ($\lambda$), logarithmic average phonon frequency $\omega_{log}$, and superconducting transition temperature ($T_c$ in K) under pressure have not been reported so far. For elastic properties, we have analyzed elastic constants, Poisson's ratio, elastic moduli, aggregate modulus, machinability index, Cauchy pressure, hardness, Kleinman parameter, and anisotropy in elastic moduli under several pressures. In addition, we have explored several thermal properties of this compound, such as the Debye temperature, melting temperature, Gruneisen parameter, lattice thermal conductivity, and sound velocities, which have not been reported yet. In this work, we have found an effect of pressure on $T_c$, DOS at the Fermi level, and strength of EPC, and logarithmic average phonon frequency $\omega_{log}$. We feel the necessity to close existing research gaps, and this drives the current work. It must be emphasized that the present work is expected to reveal  the potential of $LiMgZr_2H_{12}$ for technological applications, as well as provide guidance for experimental scientists on high $T_c$ superconductors at ambient and high pressures.

# 2 Computational Methodology

The First-principles calculations were performed within the Density Functional Theory (DFT) using the Vienna Ab initio Simulation Package (VASP) framework [**8–11**]. The projector augmented wave (PAW) method and generalized gradient approximation (GGA-PBE) functional were used to treat electron-ion interactions and exchange-correlation effects, respectively [**12–14**]. A plane wave basis was set with a kinetic energy cutoff of 600 eV to extend the electronic wave functions. Brillouin zone sampling was conducted using a Γ-centered $k$-point mesh with a reciprocal space resolution of $2\pi \times 0.02$ Å$^{-1}$ after testing. The convergence tolerance is $10^{-6}$ eV for total energy and 0.002 eV/Å for all forces.

To accurately describe the electronic properties, including the band structure and density of states (DOS), spin-orbit coupling (SOC) was explicitly accounted for. The mechanical stability and elastic constants were determined via the stress-strain method [**15**]. For the evaluation of the Fermi surface and optical properties, a dense k-point mesh of 24 × 24 × 24 was utilized.

Phonon dispersion relations were calculated within Density Functional Perturbation Theory (DFPT) [**16**] as implemented in the Quantum ESPRESSO package [**17,18**]. The electron-phonon coupling (EPC) calculations were performed using a dense 24 × 24 × 24 $k$-point grid and a 3 × 3 × 3 $q$-point grid. To ensure convergence, a plane-wave kinetic energy cutoff of 60 Ry and a Gaussian smearing of 0.02 Ry was adopted. Spin orbit coupling was not included in the phonon related calculations, as it is expected to have less importance in describing the vibrational and superconducting properties [**19,20**].

# 3 Results and discussion

## 3.1 Thermodynamic Stability and Structural Features

The thermodynamic stability of the $LiMgZr_2H_{12}$ structure can be calculated in terms of its formation energy, defined as follows:

$$\Delta E_f = \frac{E(\mathrm{LiMgZr_2H_{12}}) - E(\mathrm{Li}) - E(\mathrm{Mg}) - 2E(\mathrm{Zr}) - 12E(\mathrm{H})}{16} \tag{1}$$

where, $E(LiMgZr_2H_{12})$ denotes the total energy of the compound, and E(Li), E(Mg), E(Zr), and E(H) denote the average energies of single Li, Mg, Zr, and H atoms in the crystals, respectively. The negative value of the formation energy at all investigated pressures confirms the thermodynamic stability of the structure across the entire studied pressure range.

The gravimetric hydrogen storage capacity of the $LiMgZr_2H_{12}$ compound was evaluated based on the amount of hydrogen stored per unit mass of the material. This capacity represents how much hydrogen can be held relative to the total mass of the host system. It can be calculated using

**Equation(2)** [**21**], where $m_H$ and $m_{Host}$ denote the molar masses of hydrogen and the host material, respectively, and H/M is the ratio of hydrogen atoms to host atoms.

$$C_{\mathrm{wt\%}} = \left( \frac{\left(\frac{H}{M}\right) m_H}{m_{\mathrm{Host}} + \left(\frac{H}{M}\right) m_H} \times 100 \right) \% \quad (2)$$

The value of the gravimetric hydrogen storage capacity ($C_{wt\%}$ in wt%) for $LiMgZr_2H_{12}$ is 5.36%. With a gravimetric hydrogen storage capacity of 5.36 wt%, $LiMgZr_2H_{12}$ shows promising potentiality for the applications in portable energy systems, stationary storage, and hybrid hydrogen storage technologies.

The optimized $LiMgZr_2H_{12}$ structure is found to crystallizes in the orthorhombic structure with space group P*mmm* (No. 47) [**1**], showing a highly ordered and symmetric arrangement of atoms. The structure is mainly built from a three-dimensional zirconium-hydrogen (Zr-H) framework. In this framework, Zr atoms are surrounded by hydrogen atoms, forming coordinated polyhedral units that are connected to each other through shared hydrogen atoms, giving rise to a stable network.

Hydrogen atoms occupy different crystallographic sites which creates a complex bonding pattern throughout the structure. These hydrogen atoms act as bridges between Zr atoms and also interact with lithium and magnesium that helps to hold the structure together. Lithium and magnesium atoms are located at highly symmetric positions within the spaces of the Zr-H framework. Lithium is likely to be in a more open environment whereas magnesium tends to interact more strongly with nearby hydrogen atoms.

Overall, the structure can be viewed as a Zr-H framework that hosts Li and Mg atoms in its interstitial spaces resulting in a stable and well-organized crystal structure.

**Table 1.** Atomic coordinates of $LiMgZr_2H_{12}$ under ambient pressure.

| Compound | Atom | Fractional coordinates | | |
|---|---|---|---|---|
| | | *x* | *y* | *z* |
| $LiMgZr_2H_{12}$ | Li (1*f*) | 0.500 | 0.500 | 0.000 |
| | Mg (1*h*) | 0.500 | 0.500 | 0.500 |
| | Zr (2*q*) | 0.000 | 0.000 | 0.245 |
| | H1 (2*r*) | 0.000 | 0.500 | 0.130 |
| | H2 (2*r*) | 0.000 | 0.500 | 0.636 |
| | H3 (4*v*) | 0.500 | 0.766 | 0.252 |
| | H4 (2*i*) | 0.762 | 0.000 | 0.000 |
| | H5 (2*j*) | 0.768 | 0.000 | 0.500 |

**Table 2.** Calculated and previously obtained theoretical lattice constants (a, b, and c in Å) and equilibrium volume (V in $Å^3$) of the $LiMgZr_2H_{12}$ compound.

| Compound | P (GPa) | a | b | c | V | Ref. |
|---|---|---|---|---|---|---|
| $LiMgZr_2H_{12}$ | - | 3.7833 | 3.7853 | 7.5389 | 107.96 | [1] |
| | 0 | 3.7823 | 3.7903 | 7.5440 | 108.15 | This work |
| | 2 | 3.7567 | 3.7663 | 7.4949 | 106.04 | |
| | 4 | 3.7331 | 3.7429 | 7.4492 | 104.09 | |
| | 6 | 3.7123 | 3.7203 | 7.4069 | 102.30 | |
| | 8 | 3.6922 | 3.6995 | 7.3672 | 100.63 | |
| | 10 | 3.6737 | 3.6793 | 7.3296 | 99.07 | |

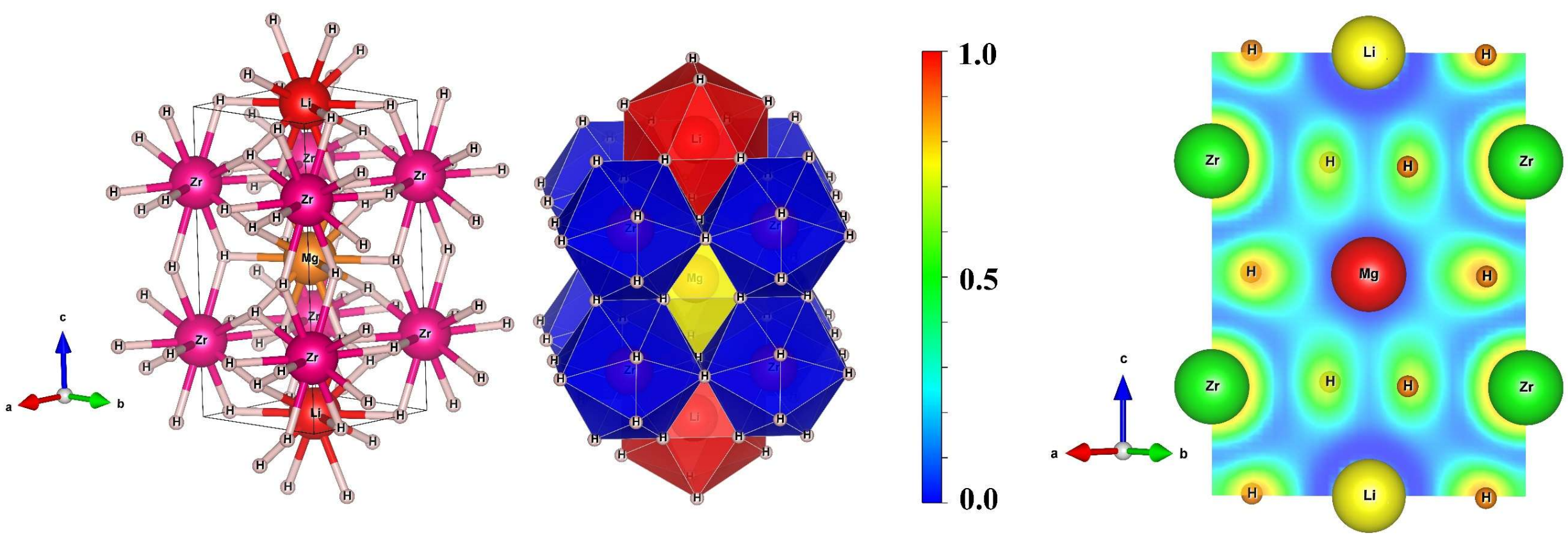

**Figure 1.** Crystal structure of the quaternary hydride $LiMgZr_2H_{12}$ with P*mmm* symmetry and electron localization function (ELF) of $LiMgZr_2H_{12}$.

To clarify the bonding characteristics in $LiMgZr_2H_{12}$, the electron localization function (ELF) and Bader charge analyses were performed. **Figure 1** presents the two-dimensional ELF map, where the weak electron localization between metal and hydrogen atoms suggests predominantly ionic interactions. This is consistent with the charge transfer from Li, Mg, and Zr atoms toward hydrogen. The ELF value between nearest-neighbor H atoms is about 0.40, indicating weak electron localization in the H-H region. Therefore, no significant H-H covalent bonding is present, suggesting that hydrogen exists mainly as monatomic species rather than $H_2$-like molecular units. The Bader charge analysis shows that Li, Mg, and Zr atoms donate approximately 0.85, 1.63, and 1.65 e, respectively, whereas each H atom gains about 0.42–0.57 e. These results confirm the dominant ionic nature of the metal-hydrogen bonding in $LiMgZr_2H_{12}$.

## 3.2 Electronic Properties

To elucidate the electronic characteristics of $LiMgZr_2H_{12}$, its electronic band structure, electronic energy density of states and Fermi surface (FS) were calculated and analyzed.

### 3.2.1 Band Structure and Density of States

To elucidate the microscopic origin of the structural stability and potential superconducting behavior in the quaternary hydride $LiMgZr_2H_{12}$, the electronic band structures were calculated along the high-symmetry directions (Γ-Z-T-Y-S-X-U-R) of the first Brillouin zone [**22,23**]. The orbital-projected electronic band structures of $LiMgZr_2H_{12}$ at ambient pressure (0 GPa) and under 10 GPa, both without and with SOC, are illustrated in **Figure 2** within an energy window of ± 6 eV around the Fermi level ($E_F$) [**24,25**]. The calculated band structures reveal that several bands cross the Fermi level in all cases, confirming the metallic nature of the compound [**26**]. This metallic behavior is consistent under both pressure conditions and is not significantly altered by the inclusion of SOC.

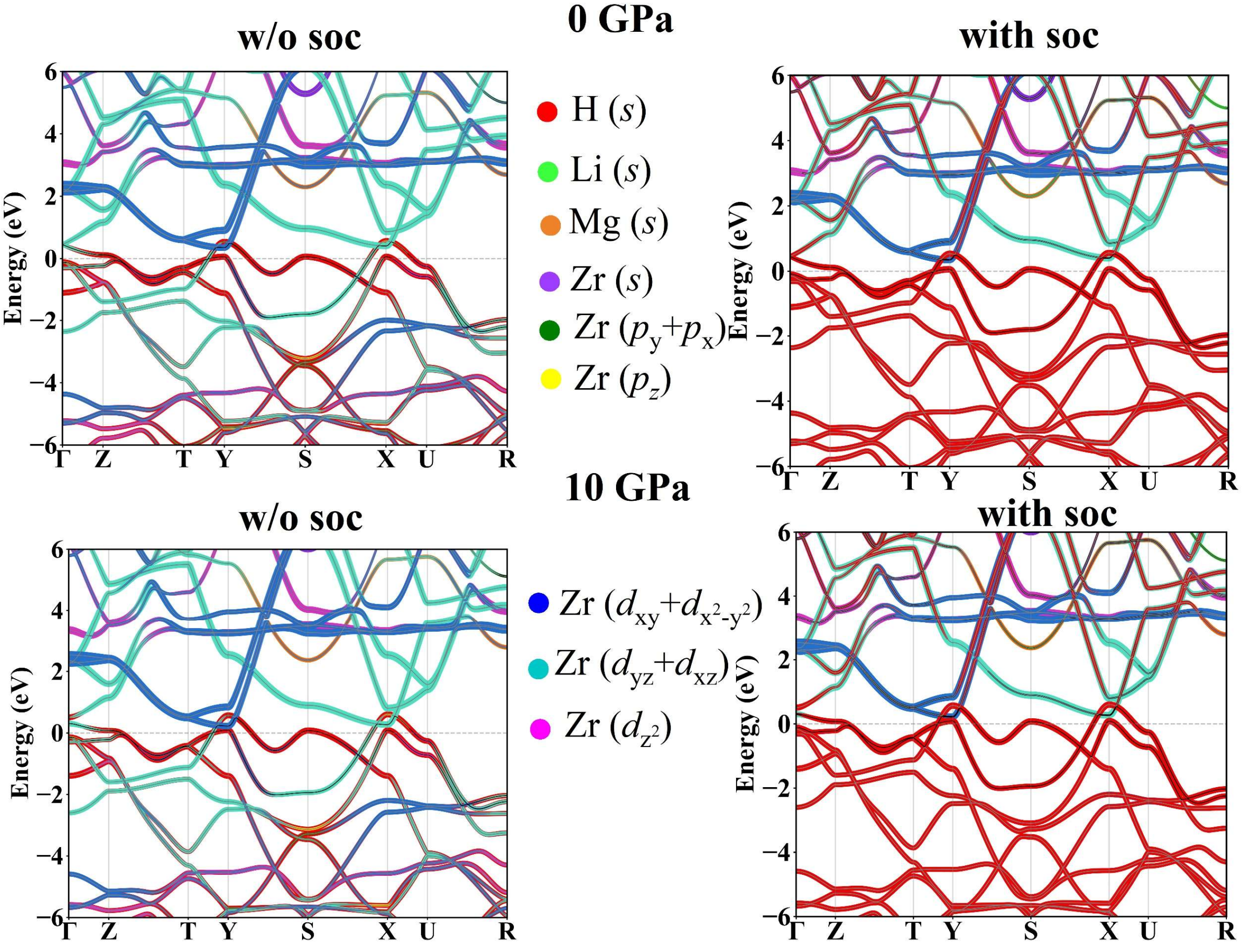

**Figure 2.** Orbital projected electronic band structures of $LiMgZr_2H_{12}$ within ± 6 eV of $E_F$ for without (w/o SOC) and with SOC at ambient pressure (0 GPa) and 10 GPa.

At 0 GPa without SOC, the bands near the Fermi level are primarily dominated by Zr-*d* orbitals particularly by the $d_{xy}$, $d_{x^2-y^2}$, $d_{yz}$, $d_{xz}$, and $d_{z^2}$ states. This indicates that the conduction properties are mainly governed by Zirconium atoms. Contributions from H-*s* states are also noticeable in the lower energy region that suggests hybridization between Zr-*d* and H-*s* orbitals. The Mg-*s* and Li-

*s* states contribute relatively weakly near the Fermi level, implying a minor role in electrical conduction.

Upon inclusion of SOC at 0 GPa, minor effects are observed especially in the Zr-*d* dominated bands due to relativistic effects [**27**]. However, the overall band topology and metallic character remain largely unchanged. This indicates that SOC has a small but non-negligible influence on the electronic structure that affects band degeneracy rather than opening a band gap.

Under an applied pressure of 10 GPa, the band dispersion becomes more pronounced which indicates increased bandwidth and enhanced orbital overlap. This is attributed to reduced interatomic distances under pressure, which strengthens the hybridization between atomic orbitals. The Zr-*d* states continue to dominate near the Fermi level and reinforcing their critical role in determining the electronic properties of the compound.

At 0 GPa with SOC additional band splitting is observed compared to the non-SOC case and the splitting becomes prominent near high-symmetry points. Despite these modifications, no band gap is observed, and the compound retains its metallic nature. The persistence of metallicity under pressure suggests robustness in its electronic structure which is important for potential superconducting behavior.

Overall, the band structure analysis demonstrates that $LiMgZr_2H_{12}$ is a pressure stable metallic system with dominant Zr-*d* orbital contributions near the Fermi level. The inclusion of SOC slightly modifies the band dispersion but does not alter the fundamental electronic characteristics. These findings are crucial for understanding the transport and superconducting properties of the compound [**28,29**].

**Figure 3** illustrates the complete evolution of the band structure under applied pressures ranging from 0 to 10 GPa. Across the entire investigated pressure range, numerous dispersive bands cross the Fermi level ($E_F$) which unambiguously confirms the intrinsic metallic character of $LiMgZr_2H_{12}$. This metallization is a fundamental prerequisite for conventional phonon-mediated superconductivity in hydrogen-rich materials [**25,28**]. As the external pressure increases from 0 GPa to 10 GPa, we observe an enhanced dispersion of the bands and a general broadening of the electronic bandwidth, which is a direct consequence of the decreased interatomic distances and increased orbital overlap.

A critical feature in the band structure is the presence of saddle points leading to van-Hove singularities (vHs), explicitly marked near the Z, T, Y and S points, situated in close proximity to the Fermi level. The proximity of a vHs to $E_F$ can significantly amplify the electronic density of states, thereby induce structural instabilities or drastically enhance the electron-phonon coupling parameter ($\lambda$), which is highly favorable for achieving a high superconducting transition temperature ($T_c$) [**30,31**]. Other pressure-sensitive band topological features are noted along the T-Y-path, where localized band shifts occur as pressure is applied.

The physical picture provided by the band structure is further corroborated by the Total and Partial Density of States (TDOS and PDOS), shown in the right panel of **Figure 3**.

The states at and immediately below $E_F$ are overwhelmingly dominated by the Zirconium (Zr) *d*-orbitals with a substantial contribution from the Hydrogen (H) *s*-orbitals. The identical peak morphologies in the energy range of -6 to 0 eV for both Zr and H signify a strong hybridization between the transition metal *d*-states and the hydrogen lattice pointing toward a mixed covalent-metallic bonding network. In contrast, the contributions from Lithium (Li) and Magnesium (Mg) at the Fermi level are notably minimal. The PDOS for these lighter elements is localized primarily at deeper energy levels or conduction band regions well above $E_F$. This implies that Li and Mg act primarily as electron donors to the Zr-H framework. This charge transfer mechanism not only stabilizes the dense hydrogenic sub-lattice but also effectively subjects the internal structure to 'chemical pre-compression' [**25,32**]. Consequently, the inclusion of these lightweight elements allows the highly symmetric, strongly coupled Zr-H network to remain dynamically stable at pressures (0-10 GPa) that are drastically lower than those required for binary super hydrides like $LaH_{10}$ or $H_3S$ [**33,34**].

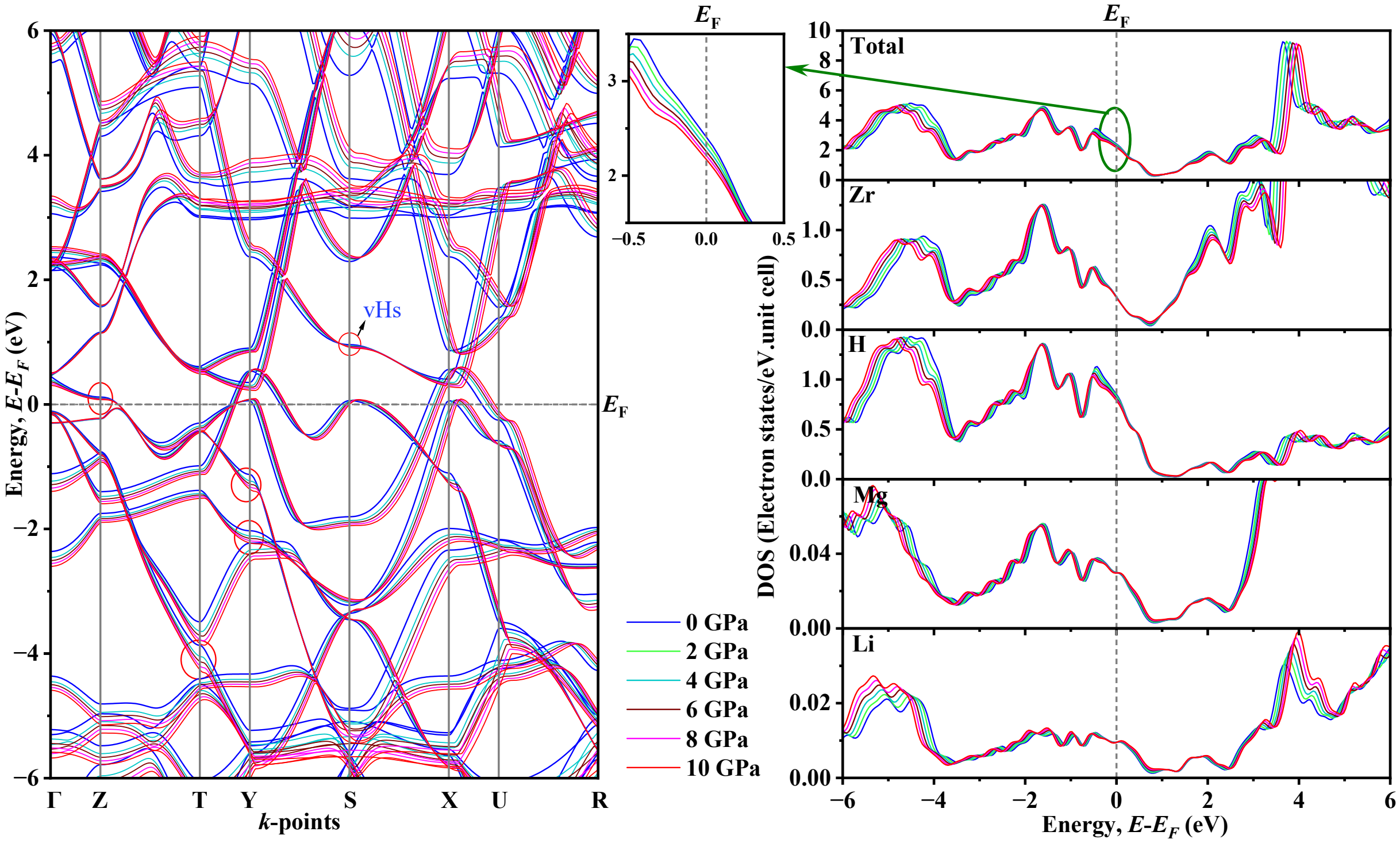

**Figure 3.** Comparison of electronic band structures and total density of states of LiMgZr2H12 along high symmetry directions in the first Brillouin zone for 0, 2, 4, 6, 8, and 10 GPa pressures.

To precisely understand the chemical bonding network, we performed orbital-resolved partial density of states (PDOS) calculations for $LiMgZr_2H_{12}$. Because the compound contains a 4*d* transition metal (Zirconium), assessing the impact of SOC on the electronic band structure near the Fermi level ($E_F$) is a critical step before proceeding to electron-phonon coupling evaluations [**35–37**]. Error! Reference source not found. illustrates the total and orbital-resolved DOS at both 0 GPa and 10 GPa.

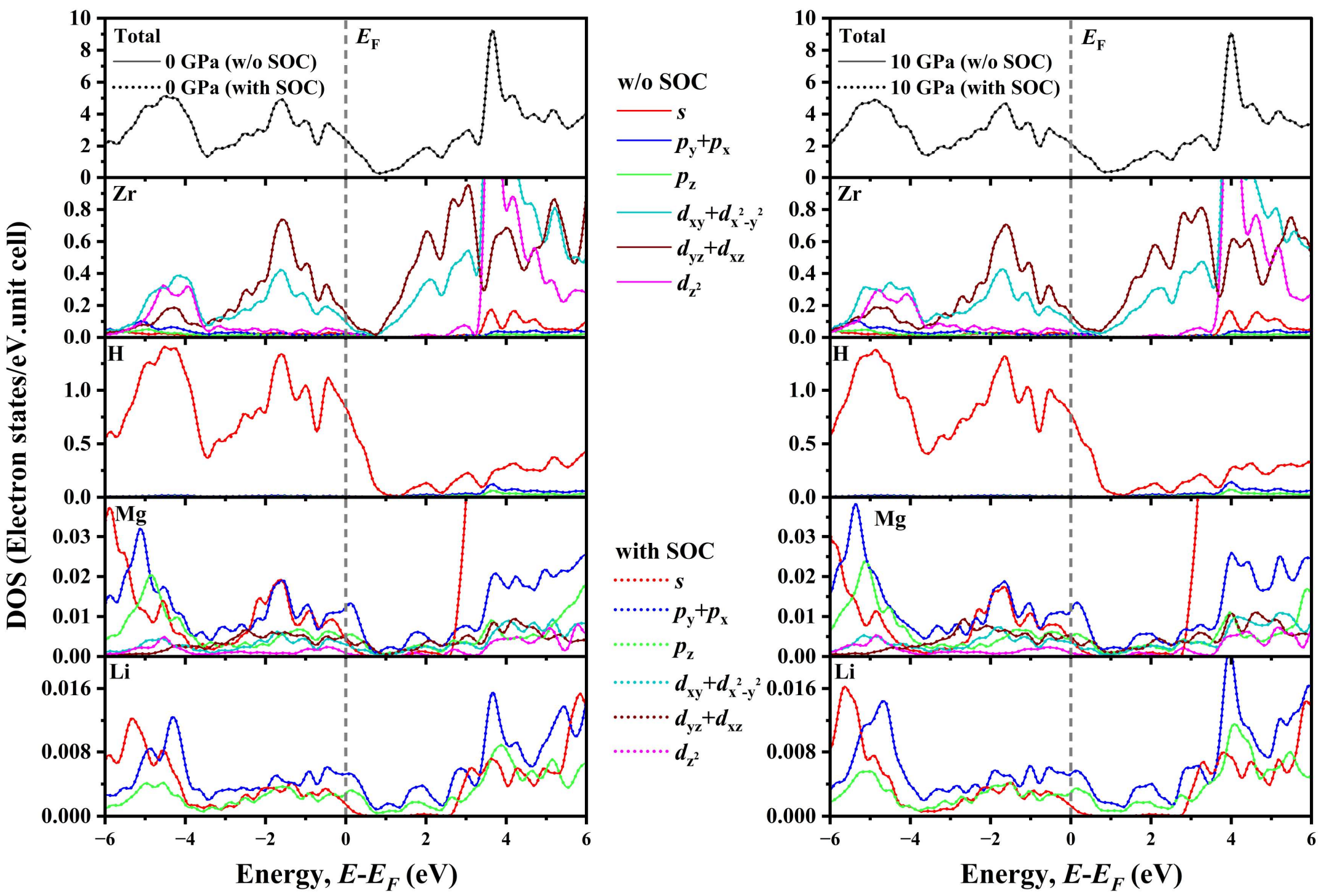

Error! Reference source not found.**.** The total and orbital-resolved DOS at both 0 GPa and 10 GPa.

As clearly demonstrated in the total DOS (top panels) and the subsequent PDOS profiles, the inclusion of SOC induces virtually no perturbation to the electronic structure. The dashed lines (with SOC) perfectly superimpose onto the solid lines (without SOC) across the entire evaluated energy range (-6 eV to +6 eV) at both ambient and elevated pressures. This indicates that the spin-orbit splitting of the Zr 4*d* bands is insufficiently large to alter the density of states at the Fermi level. Consequently, we can definitively conclude that scalar relativistic DFT calculations are thoroughly adequate for describing the electronic and potential superconducting properties of $LiMgZr_2H_{12}$, significantly reducing the computational expense for subsequent thermodynamic and phonon calculations.

The orbital-projected states reveal the microscopic origin of the compound's metallicity. The states spanning from -6 eV up to $E_F$ are overwhelmingly dominated by a strong hybridization between the Zirconium *d*-orbitals and the Hydrogen *s*-orbitals.

The bottom panels of Error! Reference source not found. highlight the critical, albeit indirect, role of the light elements, Magnesium and Lithium. The PDOS axis scales for Mg and Li are an order of magnitude smaller than those of Zr and H, and their contributions at $E_F$ are virtually zero. Their valence electrons-primarily mixed *s* and *p* character are either deeply bound in the lower valence band or pushed into the higher conduction band.

This electronic configuration implies a strong ionic character in their bonding, where Li and Mg act as nearly complete electron donors to the $Zr_2H_{12}$ anionic framework. This massive charge transfer effectively serves as a "chemical precompression" mechanism [**25,38**]. By donating electrons, Li and Mg stabilize the highly dense and hydrogen-rich sub-lattice at pressures (0-10 GPa) that are drastically lower than those required to stabilize binary transition-metal hydrides. This offers a viable pathway toward ambient-pressure high-$T_c$ superconductivity.

### 3.2.2 Fermi Surface Topology

The Fermi surface topology of $LiMgZr_2H_{12}$ was calculated throughout the Brillouin zone. As shown in **Figure 4**, four bands cross the Fermi level in $LiMgZr_2H_{12}$ confirming its metallic character. The Fermi surfaces associated with n = 1 and n = 2 form closed and smooth pockets, indicating nearly isotropic metallic carriers. In contrast, the n = 3 band produces a rhombic closed "inner-shell" like Fermi surface with relatively flat facets. These flat regions appear nearly parallel to the extended Fermi sheets generated by the n = 4 band, suggesting weak interband Fermi-surface nesting. Such nesting can enhance electronic scattering at specific wave vectors, potentially leading to phonon softening or Kohn anomalies. As a result, the electron phonon coupling may be strengthened which could contribute to the superconducting properties of $LiMgZr_2H_{12}$ [**39**]. A similar Fermi surface topology is observed at 10 GPa, indicating that the main geometrical features of the Fermi sheets are preserved under compression.

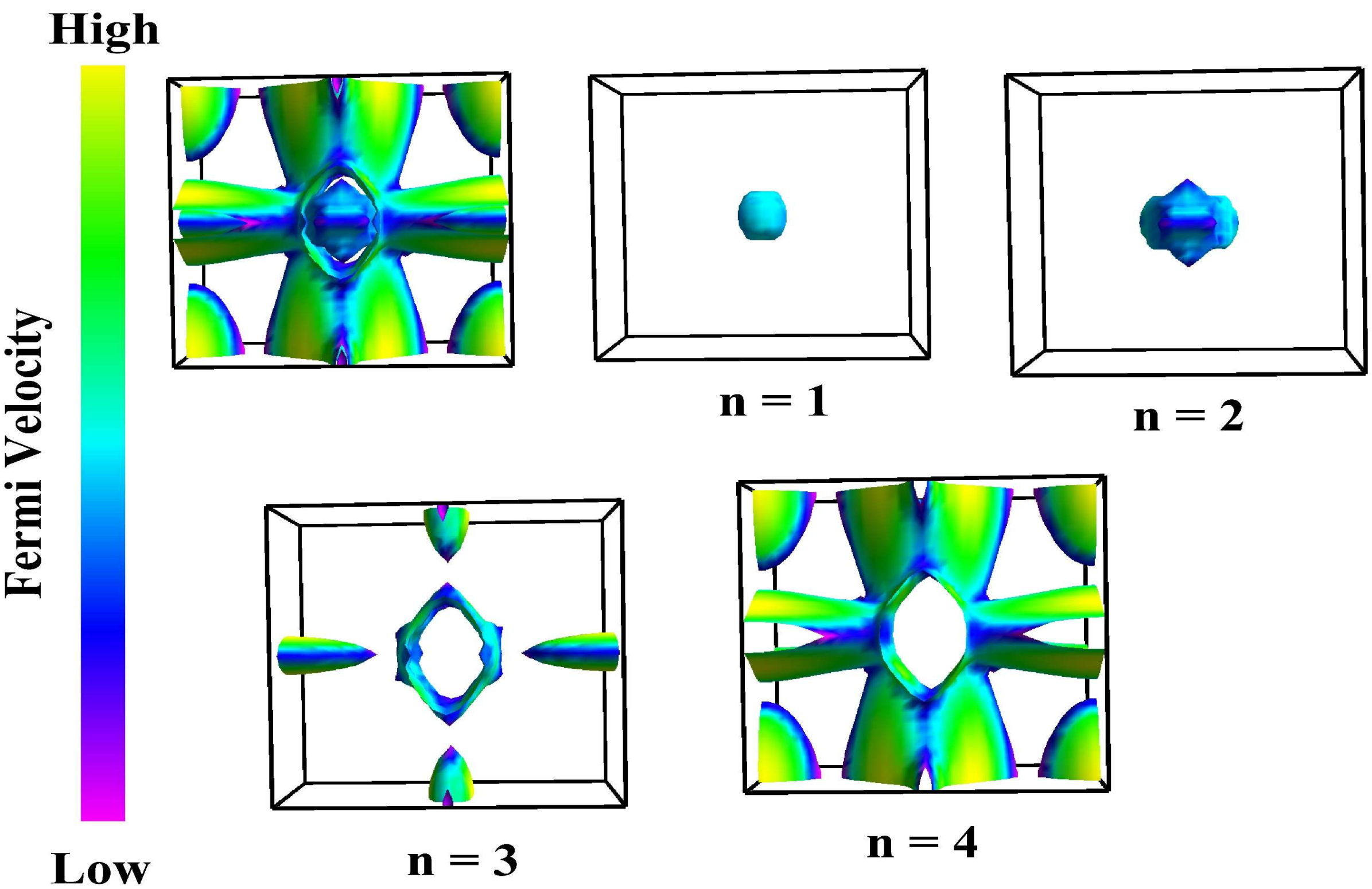

**Figure 4.** Fermi surface topologies of $LiMgZr_2H_{12}$ without SOC at 0 GPa.

## 3.3 Mechanical and Elastic Features

To understand the mechanical and dynamical properties of a material, essential from the point of view of possible applications, knowledge of its elastic constants is essential. In Voigt notation, the elastic constants of a crystal are arranged in a 6 × 6 symmetric matrix, commonly referred to as the stiffness matrix [**40**]. The crystal symmetry fixes the number of independent elements of this matrix. As our compound has an orthorhombic structure, there are nine independent components of this matrix. For a crystal to be mechanically stable, the matrix elements or the elastic constants must obey the Born stability criteria. The criteria for stability are [**41**]:

$$C_{ii} - P > 0, \quad i = 1, 4, 5, 6$$

$$(C_{11} - P)(C_{22} - P) - (C_{12} + P)^2 > 0 \qquad (3)$$

$$(C_{11} - P)(C_{22} - P)(C_{33} - P) + 2(C_{12} + P)(C_{13} + P)(C_{23} + P) - (C_{11} - P)(C_{23} + P)^2 - (C_{22} - P)(C_{13} + P)^2 - (C_{33} - P)(C_{12} + P)^2 > 0$$

$C_{11}$, $C_{22}$, $C_{33}$ are the measures of resistance to the applied stress along crystallographic directions a, b, c. In **Table 3**, calculated elastic constants for our compound $LiMgZr_2H_{12}$ are given under several pressures. These elastic constants satisfy the Born stability criteria at all pressures considered. So, the compound is mechanically stable. For this compound, $C_{11}$, $C_{22}$ and $C_{33}$ are comparable under all pressures. This indicates that the material exhibits nearly equal resistance to strain when force is applied along the a, b, and c crystallographic directions. The elastic constants such as $C_{44}$, $C_{55}$, $C_{66}$ are related to resistance to the shearing force. $C_{44}$ is a measure of resistance to shear across (100) plane with a tangential force applied in the [010] direction [**42**]. From **Table 3**, we can see that $C_{44}$, $C_{55}$, $C_{66}$ are comparable. The off diagonal elastic constants, $C_{12}$, $C_{13}$, and $C_{23}$ are also associated with the material's resistance to shear deformation. The elastic constant $C_{66}$ is comparable to $C_{44}$, which indicates equivalent stiffness for shearing along the (100) and (001) planes. For $LiMgZr_2H_{12}$, there is a noticeable effect of pressure on elastic constants as shown in **Figure 5(a)**. Except for $C_{23}$, $C_{12}$, and $C_{13}$ the value of all the elastic constants increases with increasing pressure. $C_{23}$ decreases with increasing pressure from 6 to 8 GPa, but for the rest of the pressure range, it increases. $C_{13}$ decreases with increasing pressure from 2 to 4 GPa. Therefore, applied pressure serves as an effective tool for tuning the elastic constants of this compound. The Cauchy pressures C″ was calculated under several pressures (**Table 4**). For an orthorhombic crystal, the Cauchy pressures at different planes are given by $C_1''$= ($C_{23}$−$C_{44}$) for the (100) plane, $C_2''$= ($C_{13}$−$C_{55}$) for the (010) plane, and $C_3''$ = ($C_{12}$−$C_{66}$) for the (001) plane [**43**]. To understand the nature of atomic bonding and the ductility or brittleness of a material, the Cauchy pressure is helpful. The positive value of Cauchy pressure indicates the dominance of non-directional metallic bonding and the ductile nature of the material. On the other hand, the negative value of the Cauchy pressure indicates the dominance of the directional covalent nature of bonding and brittleness of the compound [**42**]. For $LiMgZr_2H_{12}$, all the Cauchy pressures positive, and there is a noticeable effect of the external pressure on the Cauchy pressures as shown in **Table 4**. Positive Cauchy indicates the ductile nature of the compound, which is consistent with the analysis of Poisson's ratio in the following. The machinability index is a conventional parameter to identify how easy or hard it is to cut, drill, mill, and shape material during the machining process. It offers a relative measure of a material's machinability compared to a benchmark reference material. From **Table 3** and **Figure 5(d)** we can see that there is no significant effect of pressure on machinability. Moreover, the machinability of $LiMgZr_2H_{12}$ is high compared to the stainless steels and other compounds of engineering interest [**44-48**] under all pressures.

**Table 3.** Single crystal elastic constants ($\boldsymbol{C_{ij}}$ in GPa) and the machinability index ($\boldsymbol{\mu^M}$) of $LiMgZr_2H_{12}$ compound.

| P(GPa) | $C_{11}$ | $C_{12}$ | $C_{13}$ | $C_{22}$ | $C_{23}$ | $C_{33}$ | $C_{44}$ | $C_{55}$ | $C_{66}$ | $\mu^M$ |
|---|---|---|---|---|---|---|---|---|---|---|
| 0 | 143.67 | 69.80 | 70.57 | 131.91 | 82.01 | 140.02 | 49.18 | 44.88 | 51.31 | 1.943 |
| 2 | 161.63 | 84.80 | 76.20 | 152.90 | 86.38 | 142.62 | 53.58 | 50.08 | 56.45 | 1.970 |
| 4 | 172.13 | 84.39 | 73.84 | 162.24 | 89.85 | 155.91 | 57.37 | 52.37 | 60.12 | 1.909 |
| 6 | 183.35 | 90.76 | 87.12 | 174.37 | 97.63 | 169.51 | 59.44 | 56.07 | 63.40 | 2.015 |
| 8 | 190.45 | 93.96 | 87.74 | 180.46 | 96.20 | 179.11 | 62.61 | 58.80 | 64.13 | 1.962 |
| 10 | 200.92 | 95.16 | 96.81 | 191.74 | 104.21 | 193.65 | 65.53 | 61.74 | 65.36 | 1.998 |

**Table 4.** Cauchy pressures ($C''$ in GPa) for the $LiMgZr_2H_{12}$ compound under different pressures.

| P (GPa) | $C''_1$ | $C''_2$ | $C''_3$ |
|---|---|---|---|
| 0 | 32.834 | 25.694 | 18.490 |
| 2 | 32.796 | 26.121 | 28.355 |
| 4 | 32.486 | 21.478 | 24.275 |
| 6 | 38.195 | 31.044 | 27.358 |
| 8 | 33.590 | 28.941 | 29.824 |
| 10 | 38.680 | 35.070 | 29.800 |

**Table 5.** Polycrystalline bulk moduli ($\boldsymbol{B_V}$, $\boldsymbol{B_R}$, $\boldsymbol{B}$ in GPa), shear moduli ($\boldsymbol{G_V}$, $\boldsymbol{G_R}$, $\boldsymbol{G}$ in GPa), Young modulus ($\boldsymbol{Y}$ in GPa), Pugh's ratio ($\boldsymbol{G/B}$) and Poisson's ratio ($\boldsymbol{\sigma}$), for the $LiMgZr_2H_{12}$ compound in the ground state under different pressures.

| P(GPa) | $B_V$ | $B_R$ | B | $G_V$ | $G_R$ | G | Y | G/B | $\sigma$ |
|---|---|---|---|---|---|---|---|---|---|
| 0 | 95.60 | 95.50 | 95.55 | 41.96 | 39.69 | 40.82 | 107.20 | 0.427 | 0.313 |
| 2 | 105.77 | 105.39 | 105.58 | 46.01 | 43.64 | 44.82 | 117.80 | 0.425 | 0.314 |
| 4 | 109.61 | 109.38 | 109.49 | 50.12 | 48.19 | 49.11 | 128.17 | 0.449 | 0.305 |
| 6 | 119.80 | 119.75 | 119.78 | 52.56 | 50.63 | 51.60 | 135.36 | 0.431 | 0.312 |
| 8 | 122.87 | 122.51 | 122.84 | 55.25 | 53.76 | 54.50 | 142.44 | 0.444 | 0.307 |
| 10 | 130.96 | 130.95 | 130.96 | 57.87 | 56.55 | 57.21 | 149.82 | 0.437 | 0.309 |

We have calculated bulk modulus (*B*), shear modulus (*G*), Pugh's ratio (*G/B*), Young's modulus (*Y*), and Poisson's ratio ($\sigma$). These calculated results are tabulated in **Table 5**. In the calculation of elastic moduli, the Voigt approximation assumes that the strain is uniformly distributed throughout the material. It gives an upper bound on elastic moduli [**49**]. In contrast, the Reuss approximation assumes a uniform stress distribution, while allowing the strain to vary among the grains. It gives a lower bound on elastic moduli [**50**]. Hill's approximation is given by the arithmetic average of the values of the elastic modulus obtained from Voigt and Reuss methods [**51**].

The bulk modulus, $B$ describes the material's resistance to volume compression under applied pressure, whereas the shear modulus, $G$ characterizes its resistance to shape deformation. The bulk modulus is closely related to the average binding energy of the atoms in crystals. For $LiMgZr_2H_{12}$, we can see from **Table 5** that the $B$ is greater than $G$ under all pressures, which concludes that the mechanical failure of these compounds is controlled by applied shear component. Young's modulus ($Y$) is a measure of resistance to longitudinal strain. Young's modulus and Poisson's ratio were determined from the calculated bulk and shear moduli using standard isotropic elastic relations [**52**]. From **Table 5**, it is seen that the value of Young's modulus is high and comparable to that of pure copper under all pressures, which indicates the hard nature of the material.

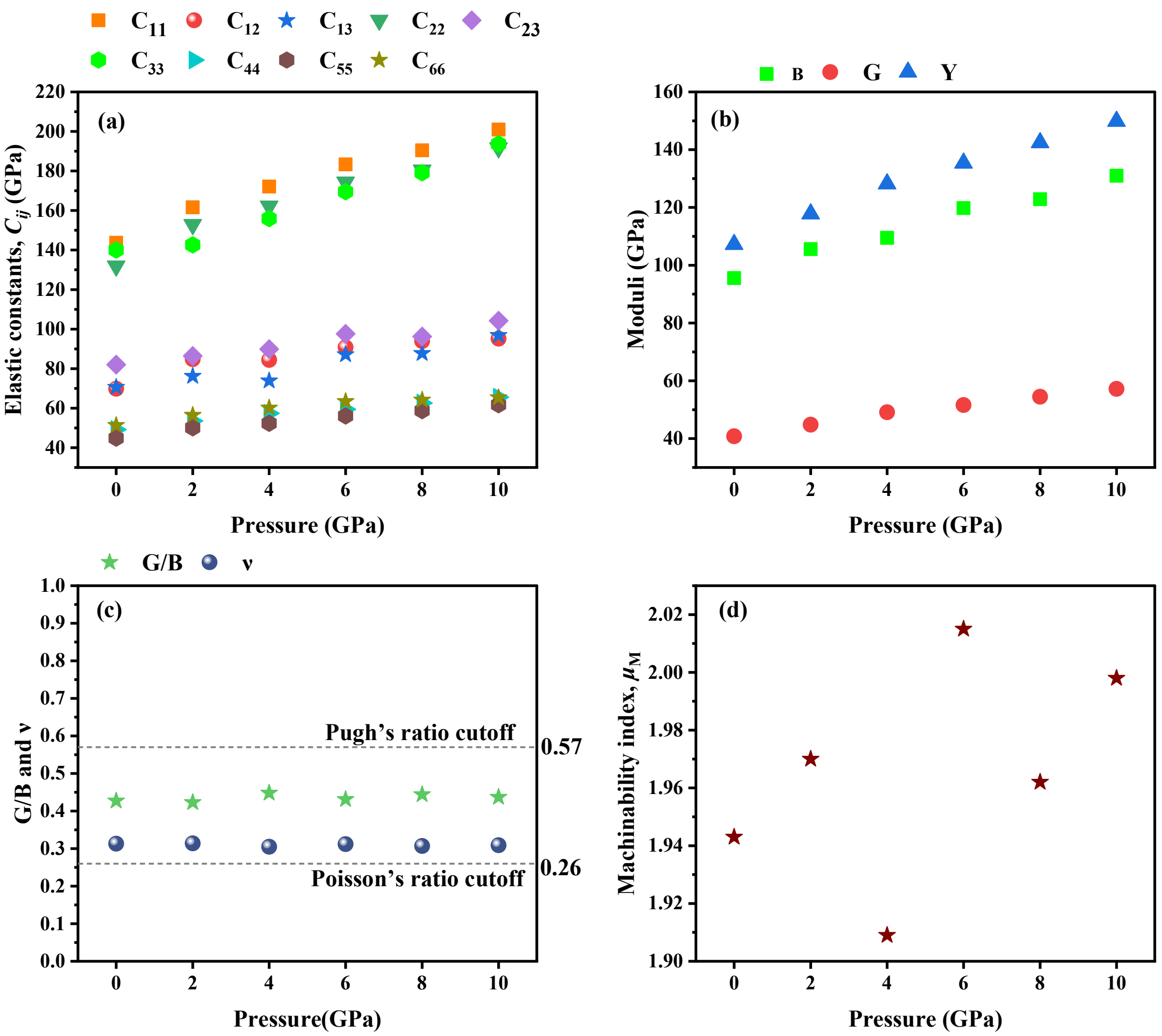

**Figure 5.** (a) Elastic constants (b) Calculated elastic moduli (c) Calculated Poisson's ratio and Pugh's ratio (d) Calculated machinability index under 0, 2, 4, 6, 8, 10 GPa.

Material's stability under the shearing stress is related to Poisson's ratio, $\sigma$. By this quantity, the brittle, ductile nature and chemical bonding of materials can be understood. Poisson's ratio is defined by the ratio of transverse strain to longitudinal strain for a definite tensile stress [**53**].

The material is considered to be ductile if its Poisson's ratio is greater than 0.26; otherwise, the material will be brittle [**53,54**]. From **Table 5**, it is seen that Poisson's ratio is greater than 0.26 under all pressures, which indicates the ductile nature of the compound. The bonding nature of the material can be gauged from Poisson's ratio. If it is within 0.25 to 0.50, then the material will be stabilized by the central force. If it is outside the range, then the material will be stabilized by a non-central force. From **Table 5**, it is obvious that the compound is predicted to be stabilized by a central force.

The Pugh's ratio, defined by $(G/B)$ [**55**], is another parameter to estimate the brittle and ductile nature of a compound. If the Pugh ratio is less than 0.57, then the material is ductile; otherwise, it is brittle [**56**]. For $LiMgZr_2H_{12}$, the calculated Pugh's ratio remains below 0.57 at all pressures considered, indicating that the material exhibits ductile nature. This result is also consistent with the behavior predicted from Poisson's ratio. From **Figure 5(b) and (c)**, we can see that the Young's modulus, bulk modulus, and shear modulus increase with increasing pressure, but there is no significant effect of pressure on Poisson's ratio and Pugh's ratio.

**Table 6.** Calculated hardness (GPa) based on elastic moduli and Poisson's ratio for the compound under different pressures.

| P (GPa) | $H_{micro}$ | $(H_V)_{Chen}$ | $(H_V)_{Tian}$ | $(H_V)_{Teter}$ | $(H_V)_{Mazhnik}$ |
|---|---|---|---|---|---|
| 0 | 5.089 | 3.476 | 4.835 | 6.164 | 5.770 |
| 2 | 5.558 | 3.789 | 5.128 | 6.768 | 6.351 |
| 4 | 6.384 | 4.637 | 5.824 | 7.415 | 6.801 |
| 6 | 6.465 | 4.499 | 5.761 | 7.791 | 7.274 |
| 8 | 7.011 | 5.016 | 6.193 | 8.230 | 7.587 |
| 10 | 7.287 | 5.098 | 6.297 | 8.639 | 8.009 |

Another important aspect of our research involves calculating the microhardness. Microhardness has many aspects to consider when discussing the physical characteristics of solids, one of which includes how all of the following items will relate to each other: ionic potential or ionicity, bond length between atoms, and density of valence electrons. Furthermore, physical defects in the crystalline structure (for example, dislocations and other point defects) will have a large effect on the measured microhardness value of the material. Materials that have a high value of hardness parameters are not easy to machine. But they are comparatively more beneficial as cutting tools and wear-resisting coatings [**42**].

The microhardness was estimated from Young's modulus and Poisson's ratio using the formalism developed in Ref. [**57**]. For this study, from **Table 6** we see the value of $H_{micro}$ varies in the range from 5.089 to 7.287 GPa for pressure from 0 to 10 GPa. This indicates the moderately hard nature of the material. Moreover, it has also become obvious that external pressure can be a useful tool to adjust the hardness of the material (**Figure 6**). This gives an indication of a material's resistance to plastic deformation as well [**42**]. In **Table 6**, the displayed Vickers hardness $(H_V)$ values were evaluated using several established models [30-33]. In **Figure 6**, we can observe that there is a significant effect of external pressure on the value of Vickers hardness for all the models.

Microcracks formation in a crystal and the crystal's durability is heavily linked with elastic anisotropy in crystals Information on elastic anisotropy is crucial for materials engineering. Elastic anisotropy in crystals is closely related to the development of microcracks and the endurance of crystals [**42**]. A criterion for determining the degree of anisotropy in bonding strength in various atomic planes is provided by the shear anisotropy factors ($A_i$). For an elastically isotropic crystal, the value of the shear anisotropy factor is one. If the value of shear anisotropy is greater than or less than one, then there is an anisotropy in response to shear along different crystal planes [**57**]. The expressions for the shear anisotropy factor for $A_1(100)$ shear planes between [011] and [010], the $A_2(010)$ shear planes between [101] and [001], and the $A_3(001)$ shear planes between [110] and [010] are provided by Refs. [**59,60**].

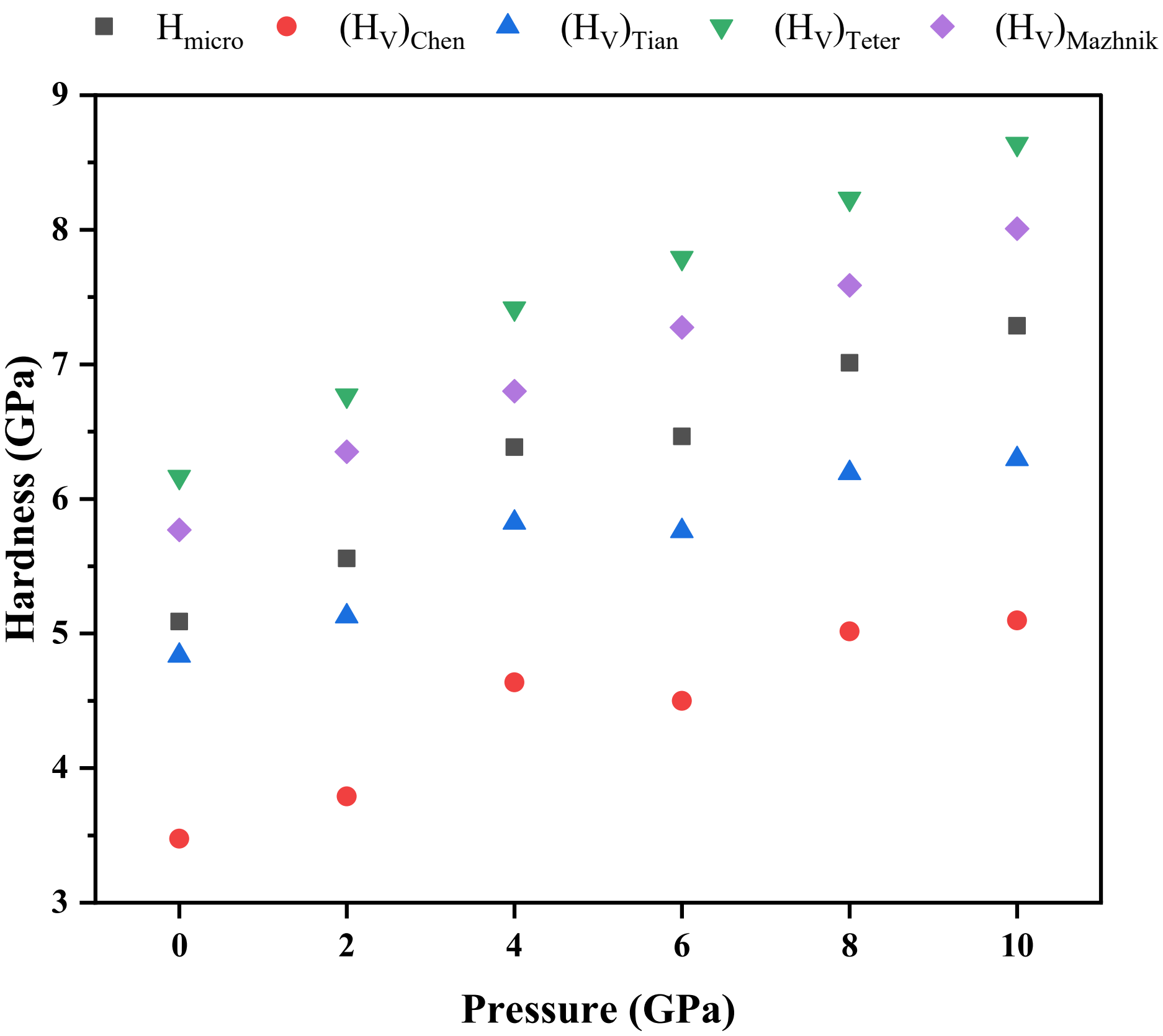

**Figure 6.** Calculated hardness parameters under several pressures.

For $LiMgZr_2H_{12}$, from **Table 7,** we can see that the shear anisotropy factors $A_1$, $A_2$, and $A_3$ are all slightly greater than one at all pressures. This indicates there is anisotropy in response to shearing stress along different crystal planes.

**Table 7.** Shear anisotropy factors ($A_1$, $A_2$ and $A_3$), anisotropy in shear ($A_G$) and anisotropy in compressibility ($A_B$), and the universal anisotropy index ($A^U$) for $LiMgZr_2H_{12}$ compound.

| P (GPa) | $A_1$ | $A_2$ | $A_3$ | $A_B$ (%) | $A_G$ (%) | $A^U$ |
|---|---|---|---|---|---|---|
| 0 | 1.380 | 1.664 | 1.509 | 0.052 | 2.781 | 0.287 |
| 2 | 1.411 | 1.632 | 1.558 | 0.179 | 2.646 | 0.275 |
| 4 | 1.272 | 1.513 | 1.452 | 0.107 | 1.960 | 0.202 |
| 6 | 1.331 | 1.509 | 1.439 | 0.023 | 1.868 | 0.191 |
| 8 | 1.290 | 1.407 | 1.402 | 0.147 | 1.369 | 0.142 |
| 10 | 1.304 | 1.395 | 1.292 | 0.004 | 1.154 | 0.117 |

The anisotropy in compressibility and shear is given by $A_B$ and $A_G$ and are calculated following Ref. [**61,62**]. For an isotropic crystal, the values of $A_B$ and $A_G$ are zero. If these values are other than zero and positive, then the crystal has anisotropy [**62**]. From the **Table 7,** it is seen that the both $A_B$ and $A_G$ have values greater than zero under all pressures. So, there is anisotropy in both volume and shear strain. Moreover, we can see that the values of $A_G$ are much greater than $A_B$ under all pressures. This result suggests that $LiMgZr_2H_{12}$ shows stronger anisotropy in shear deformation compared to volumetric deformation. Moreover, we can see from the **Table 7** at pressures 0, 6, and 10 GPa, the value of $A_B$ is very close to zero, and the values of $A_G$ are much greater than $A_B$, which implies that under those pressures the compound is highly isotropic for volume deformation and highly anisotropic for shear deformation. Universal anisotropy index ($A^U$) is also calculated following Ref. [**63**].

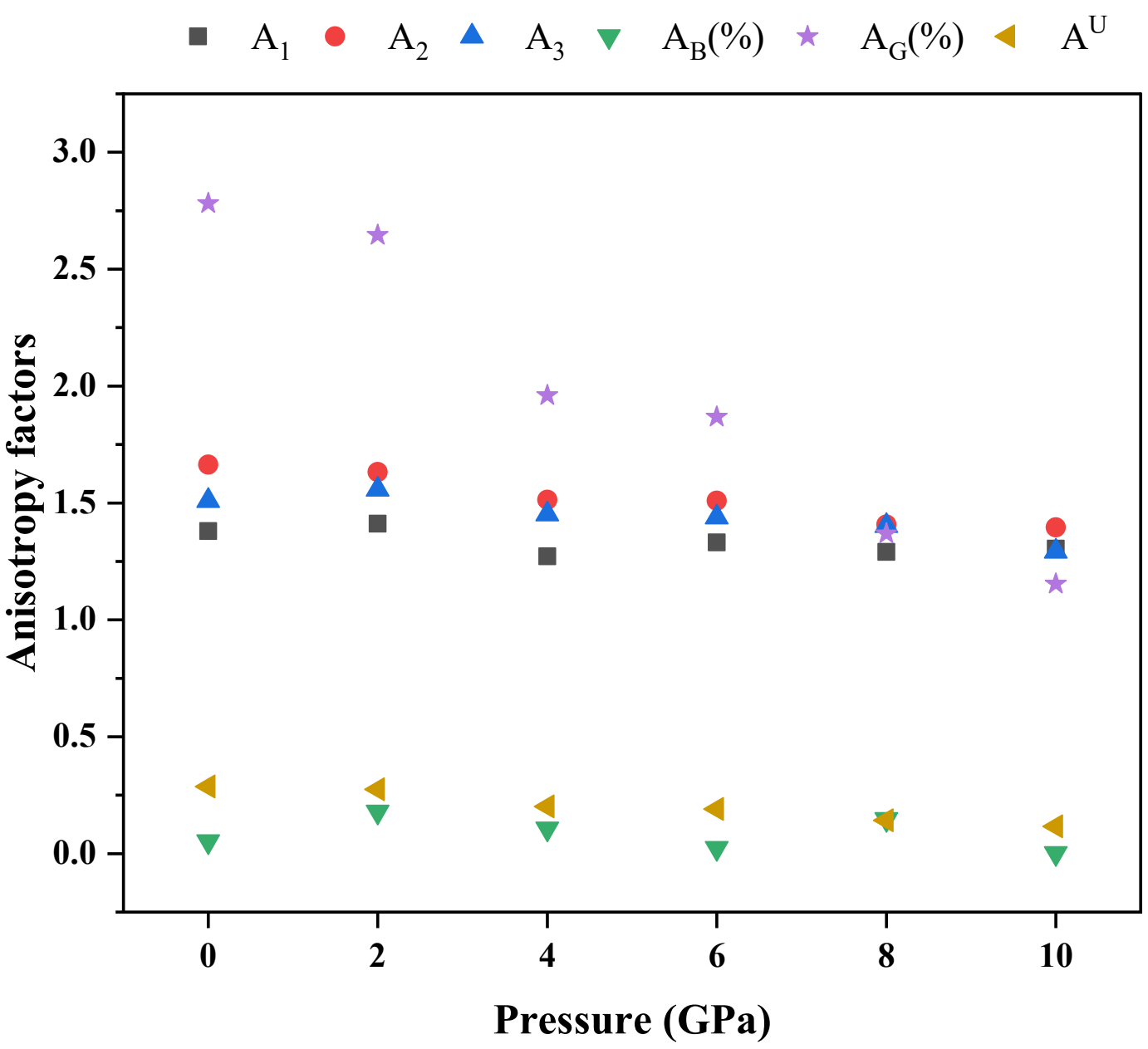

**Figure 7.** Anisotropy factors $A_1$, $A_2$, $A_3$, $A_B$, $A_G$, $A_U$ under several pressures for $LiMgZr_2H_{12}$.

Universal anisotropy index ($A^U$) is a single measure of anisotropy that is independent of crystal symmetry. It accounts for both shear and bulk anisotropies [61,63]. For an isotropic crystal, the Universal anisotropy index $A^U$ is zero, and if there is any anisotropy, then $A^U$ is greater than zero. In the **Table 7**, it is seen that for $LiMgZr_2H_{12}$, the Universal anisotropy factor is greater than zero. At zero pressure, it has the highest value of 0.287; with increasing pressure, it decreases, and at 10 GPa, it has the value 0.117. So, this also confirms the anisotropic nature of the compound under pressure. From **Figure 7** we can see that there is a significant effect of pressure on anisotropy in shear $A_G$. With increasing pressure, it decreases. There is a slight effect of pressure on the shear anisotropy factors $A_1$, $A_2$, and $A_3$. The pressure effect on $A_B$ and $A^U$ is also noticeable.

The Kleinman parameter ($\zeta$) is related to the bonding property of the material, which describes the relative positions of the cation and anion sublattices for distortions in which volume is conserved. Moreover, this parameter is independent of the crystal symmetry. Kleinman parameter ($\zeta$), gauges a compound's resistance to stretching and bending distortions. The range of values for $\zeta$ is from 0 to 1 [**64**]. The lower limit of $\zeta$ indicates that the major contribution to resistance is from bond stretching. On the other hand, the upper limit of $\zeta$ indicates that the

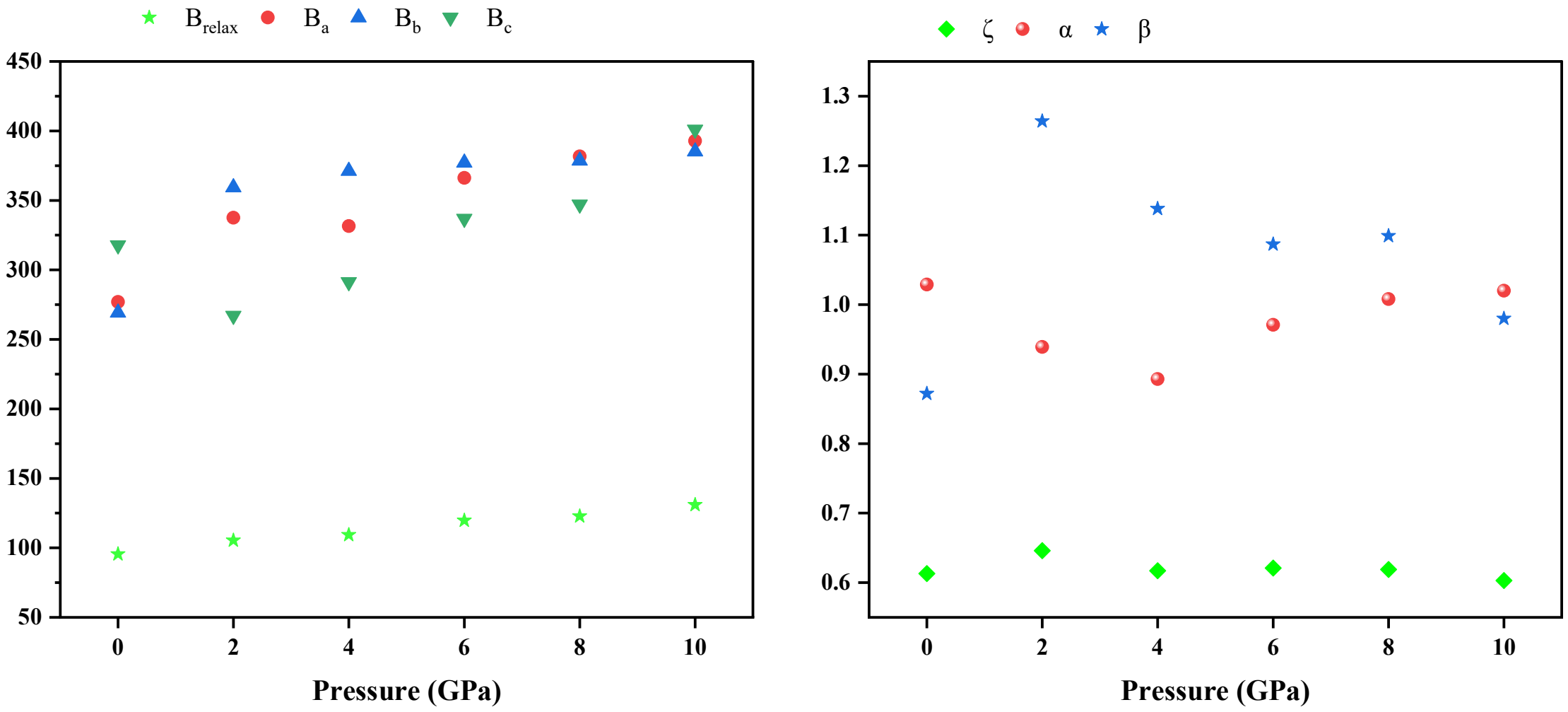

**Figure 8.** Lower bound of bulk modulus, bulk modulus along the crystallographic axes a, b, c ($B_a$, $B_b$, $B_c$ in GPa), Kleinman parameter, α, β for several pressures.

major contribution to resistance is from bond bending in response to external load. The Kleinman parameter was calculated following Ref. [**64**]. From

**Table** 8, we see that the Kleinman parameter of $LiMgZr_2H_{12}$ lies between 0.603 and 0.646 at different pressures. So, the bond-bending contributions are the main source of mechanical strengths in $LiMgZr_2H_{12}$.

**Table 8.** Calculated Kleinman parameter ($\boldsymbol{\zeta}$), bulk modulus ($\boldsymbol{B_{relax}}$ in GPa), bulk modulus along a-, b- and c-axis ($\boldsymbol{B_a}$, $\boldsymbol{B_b}$, $\boldsymbol{B_c}$ in GPa), $\alpha$, $\beta$ and Aggregate modulus ($\Lambda$) of $LiMgZr_2H_{12}$ compound under different pressures.

| P (GPa) | $\zeta$ | $B_{relax}$ | $B_a$ | $B_b$ | $B_c$ | $\alpha$ | $\beta$ | $\Lambda$ |
|---|---|---|---|---|---|---|---|---|
| 0 | 0.613 | 95.501 | 277.00 | 269.25 | 317.77 | 1.029 | 0.872 | 803.416 |
| 2 | 0.646 | 105.391 | 337.59 | 359.41 | 267.10 | 0.939 | 1.264 | 1081.361 |
| 4 | 0.617 | 109.375 | 331.55 | 371.26 | 291.27 | 0.893 | 1.138 | 1005.031 |
| 6 | 0.621 | 119.745 | 366.16 | 377.17 | 336.85 | 0.971 | 1.087 | 1119.628 |
| 8 | 0.619 | 122.810 | 381.61 | 378.60 | 347.12 | 1.008 | 1.099 | 1185.801 |
| 10 | 0.603 | 130.955 | 392.83 | 385.14 | 400.95 | 1.020 | 0.980 | 1178.389 |

The isotropic bulk modulus $B_{relax}$, the directional bulk modulus (represented by $B_a$, $B_b$, and $B_c$ along the a-, b-, and c-axis, respectively) and $\alpha$ and $\beta$, which are the relative change of the b and c axis as a function of deformation of the a-axis, for this compound were also calculated following Ref. [**59**].

The value obtained by the Reuss approximation is almost the same as the single-crystal isotropic bulk modulus $B_{relax}$ [**59**]. From **Table 5 and**

**Table** 8, we can see that the Reuss approximation of bulk modulus agrees well with the $B_{relax}$ under all pressures. From

**Table** 8, it is seen that at zero pressure bulk modulus along the b-axis, $B_b$, has the lowest value, which is 269.25 GPa, compared to the bulk modulus along the a-axis, $B_a$, and along the c-axis, $B_c$. This indicates the higher compressibility of the crystal along the b-axis. For pressures 2, 4, 6, and 8 GPa, the values of $B_c$ are 267.10, 291.27, 336.85, 347.12 GPa, respectively, and these are the lowest values in comparison to the other two directional bulk moduli. Therefore, under these pressures, the compressibility along c-axis is the highest. For 10 GPa, the $B_b$ has the lowest value in comparison to $B_a$ and $B_c$, which indicates the highest compressibility along the b-axis. From **Figure 8**, we can see that there is a significant effect of pressure on $B_a$, $B_b$, and $B_c$, but there is a small effect of pressure on $B_{relax}$. Moreover, from **Figure 8**, we see a noticeable effect of pressure on $\beta$ from 0 to 2 GPa, its value changes from 0.872 to 1.264; with further increase in pressure, it decreases and becomes 0.980 at 10 GPa. There is also a noticeable effect of pressure on $\alpha$. Finally, the Kleinman parameter $\zeta$ is found to be fairly insensitive to pressure.

## 3.4 Thermo-physical Properties

Thermophysical properties characterizes how a material response to the application of heat. The pressure-dependent thermal properties of $LiZr_2MgH_{12}$ are investigated using theoretical approaches, as experimental studies under extreme conditions remain challenging. The thermal behavior is characterized in terms of the Debye temperature, melting temperature, and phonon thermal conductivity. The corresponding expressions used for these estimations are adopted from previous reports [**65–67**].

### 3.4.1 Debye Temperature

The Debye temperature ($\theta_D$) is a fundamental parameter that provides insight into the lattice dynamics and thermodynamic behavior of crystalline solids. As it is directly related to the maximum phonon frequency and thus reflects interatomic bonding strength. Consequently, $\theta_D$ is closely linked to key properties such as specific heat, melting temperature, elastic stiffness, and electron-phonon interactions. The pressure dependence of $\theta_D$ [**Figure 9(a)**] is evaluated using the Anderson model which reliably incorporates elastic properties and density. At ambient conditions, the calculated $\theta_D$ is comparatively low across all the pressures. The Debye temperature ($\theta_D$) shows a steady increase with pressure. This is due to the lattice stiffening resulting from reduced interatomic distances and enhanced bonding strength which in turn suggests improved thermodynamic stability and a rise in melting temperature. The Debye temperature of $LiMgZr_2H_{12}$ is comparable to those of high-$T_c$ cuprates [**68**].

### 3.4.2 Melting temperature

The most significant thermophysical parameter for determining the application of materials in high temperature environments is the melting temperature ($T_m$). This property is especially relevant for materials to be used in energy storage and hydrogen related technologies. **Figure 9(b)** illustrates the pressure dependence of $T_m$ for $LiMgZr_2H_{12}$. The melting temperature ($T_m$) of $LiMgZr_2H_{12}$ is approximately 1400 K at ambient pressure and gradually increases with pressure. This behavior indicates that the thermal stability and structural robustness of $LiMgZr_2H_{12}$ enhances under compression.

At ambient conditions, the melting point is lower than that of conventional refractory materials. However, the positive pressure coefficient of $T_m$ suggests that $LiMgZr_2H_{12}$ can retain its structural integrity in relatively high-temperature and high-pressure environments. This behavior is particularly relevant for hydrogen storage systems where thermal stability plays a vital role in performance and safety. Furthermore, the variation of $T_m$ with pressure provides useful insights into the compressibility and the possibilities of pressure induced phase transitions in the system.

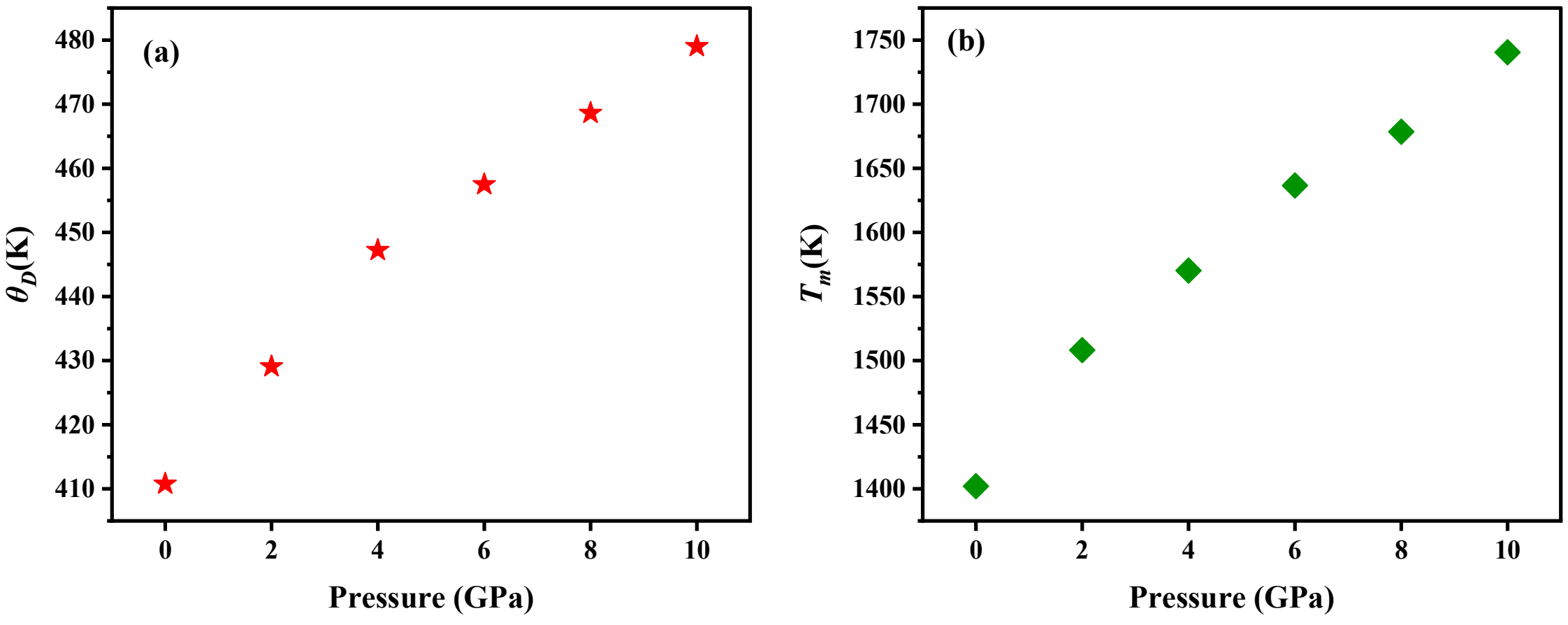

**Figure 9.** Variation of (a) Debye temperature ($\theta_D$) and (b) melting temperature ($T_m$) of $LiMgZr_2H_{12}$ with pressure.

### 3.4.3 Phonon thermal conductivity

The phonon thermal conductivity ($\kappa_{Ph}$) plays a crucial role in wide range of technological applications like thermal barrier coatings (TBCs), thermal management in electronic applications and thermoelectric systems. Materials with the high value of ($\kappa_{Ph}$) are essential for efficient heat dissipation in micro and nano-electronic applications. In contrast, a low value of ($\kappa_{Ph}$) is highly desirable for thermoelectric energy conversion and thermal insulation. The lattice thermal conductivity is strongly influenced by the atomic masses of the constituent elements and bonding strength. **Figure 10** shows the variation of ($\kappa_{Ph}$) for $LiMgZr_2H_{12}$ under different pressures and temperatures. At ambient conditions, $LiMgZr_2H_{12}$ exhibits a relatively low thermal conductivity of 0.87 W/m.K. With increasing pressure up to 10 GPa, $\kappa_{Ph}$ shows moderate pressure dependence. The gradual increment of $\kappa_{Ph}$ with pressure indicates the heat transport enhancement of $LiMgZr_2H_{12}$ under compression. This behavior can be attributed to competing effects between the reduction in phonon lifetimes and the enhancement of phonon group velocities with increasing pressure. Furthermore, the value of $\kappa_{Ph}$ gradually decreases with increasing temperature. This behavior is primarily due to enhanced phonon-phonon scattering arising from temperature induced anharmonic lattice vibrations. This trend is consistent with typical phonon dominated heat transport in complex hydride systems.

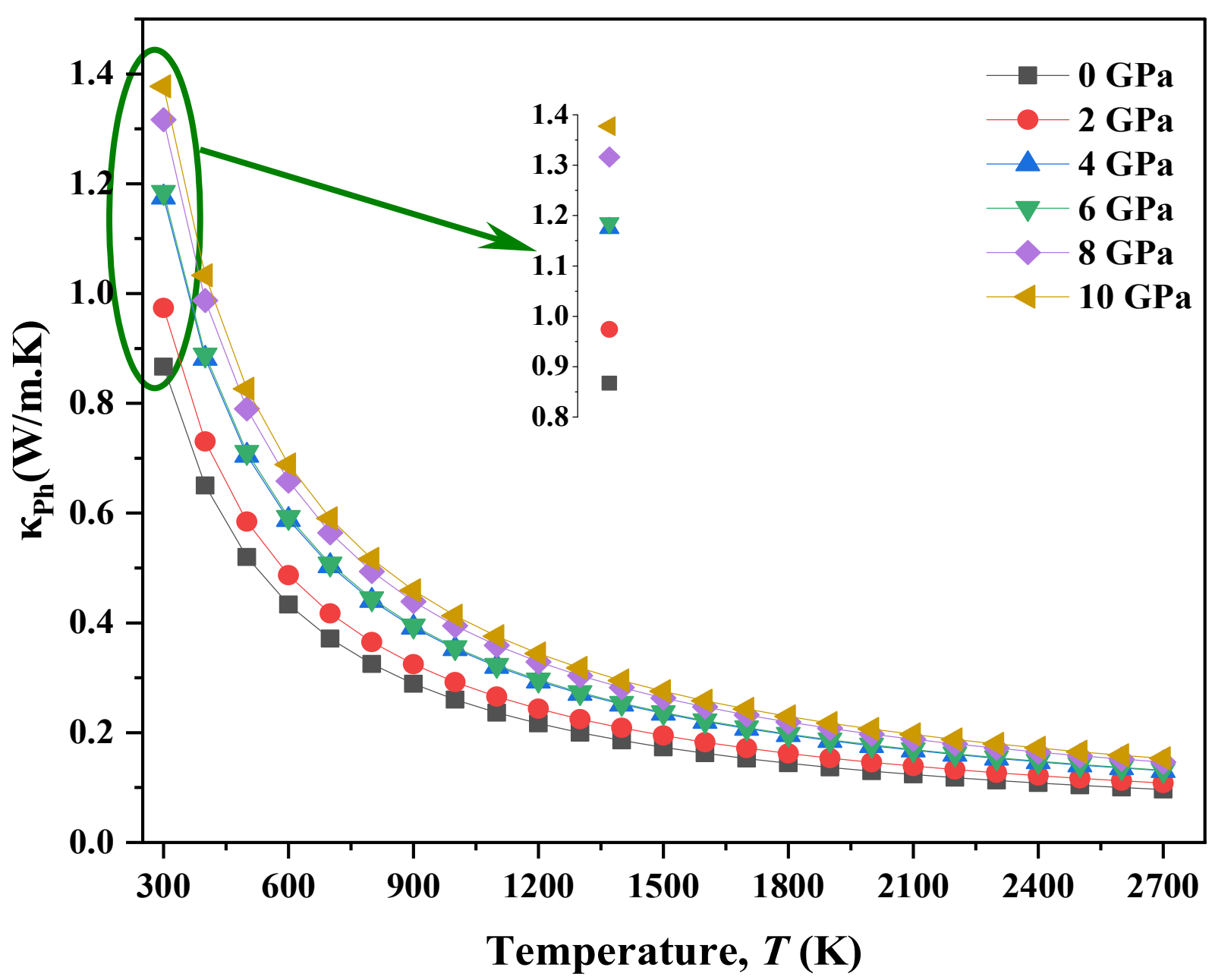

**Figure 10.** Variation of phonon thermal conductivity of $LiMgZr_2H_{12}$ as a function of pressure and temperature.

## 3.5 Optical properties

Understanding energy/frequency-dependent optical properties is necessary to forecast a material's response to incident electromagnetic radiation. Furthermore, it is a useful tool for investigating the electronic band structure, impurity level states, excitons, particular magnetic excitations, lattice vibrations, and localized defects [**69**]. From the viewpoint of optoelectronic applications, the response of materials to visible light is very important. The study of optical properties of solids has gained considerable attention because of their wide use in modern technologies such as display devices, sensors, lasers, photodetectors, photonics, solar cells, and photo-electrodes. Moreover, optical anisotropy must be considered, as many optical technologies, such as displays, 3D movie screens, polarizers, and wave plates, are based on this property. Different energy- or frequency-dependent optical properties, such as the dielectric function, loss function, refractive index, optical conductivity, absorption coefficient, and reflectivity, fully describe how a material responds to incident light [**70**].

### 3.5.1 Dielectric Function

The electric polarization of the material is described by the real (Re) component of the dielectric function, $\epsilon_1$. Conversely, the imaginary (Im) part of the dielectric function, $\epsilon_2$, represents the dielectric loss in the optical system [**70,71**]. In **Figure 11(a) and (b)**, for $LiMgZr_2H_{12}$, the real and imaginary parts of the dielectric functions are plotted respectively for [100],[010], and [001] polarization directions of the electric field under 0 and 10 GPa. In the region from 2 eV to 3 eV, it is seen that the real part of the dielectric constant $\epsilon_1$ crosses zero from a negative value for all directions of polarizations. This is an indication of the metallic nature of the material. At about 18.1 eV and 19.1 eV photon energies, the real part of the dielectric constant crosses zero from negative values for 0 and 10 GPa, respectively. The imaginary part of the dielectric constant $\epsilon_2$ becomes zero at about 20 eV. Therefore, beyond this energy, there is no dielectric loss. There is no noticeable anisotropy or effect of pressure in both the real and imaginary parts of the dielectric constant for this compound.

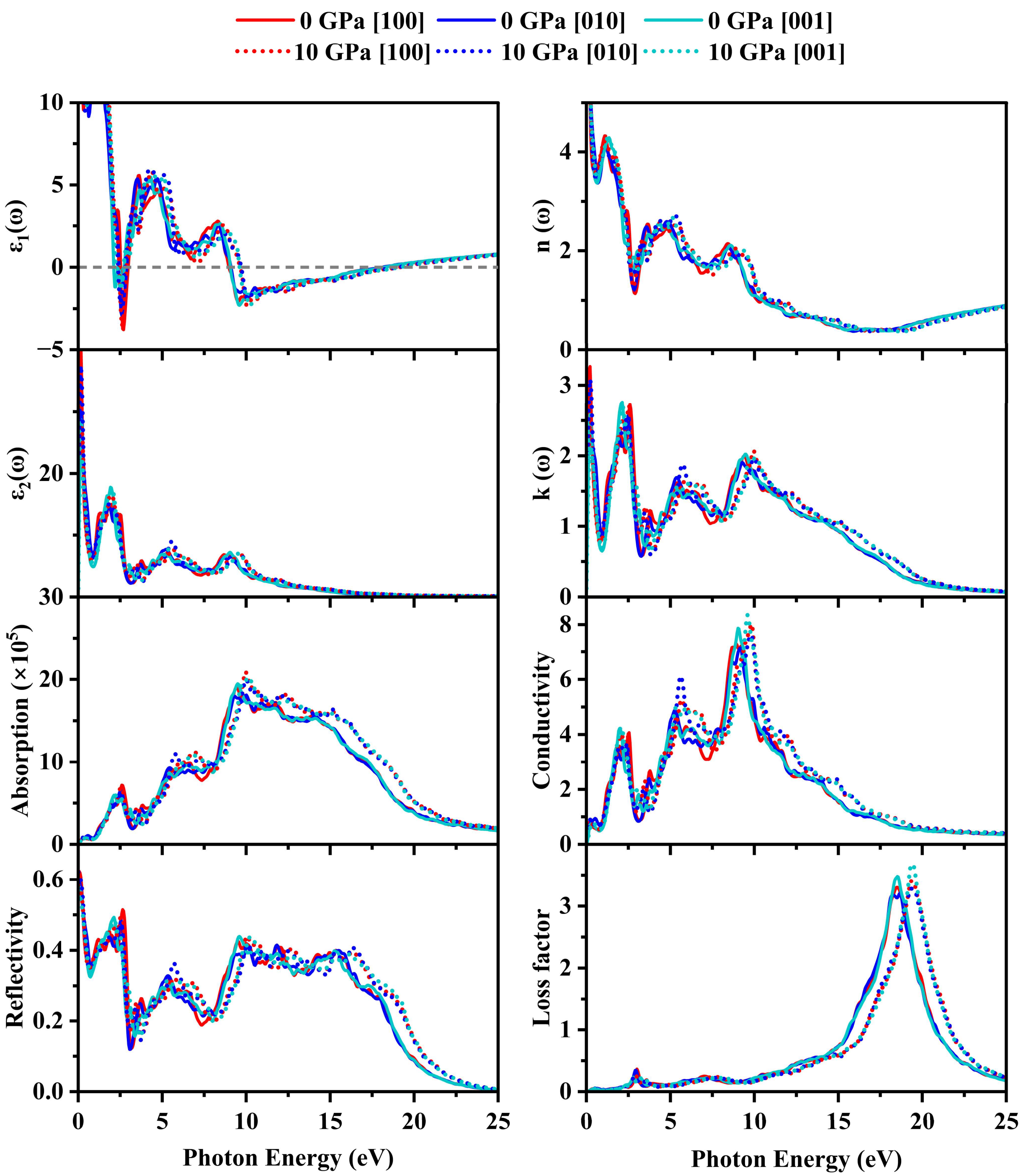

**Figure 11.** The energy (or, equivalently, frequency) dependent (a) real part of dielectric function (b) imaginary part of dielectric function (c) absorption coefficient (d) reflectivity (e) real part of refractive index (f) imaginary part of refractive index (g) optical conductivity and (h) loss function of $LiMgZr_2H_{12}$ with electric field polarization vectors along [100], [010] and [001] directions at 0 GPa and 10 GPa pressure.

### 3.5.2 Absorption Coefficient

A material's ability to absorb incident electromagnetic radiation can be predicted using the absorption coefficient ($\alpha$). The electronic nature of the material can be known from the energy-dependent absorption spectra. The energy-dependent absorption spectra of $LiMgZr_2H_{12}$ for [100], [010], and [001] polarization directions for 0 and 10 GPa are plotted in **Figure 11(c)**. The non-zero value of the absorption coefficient, very close to zero energy, indicates the metallic nature of the material, which is consistent with the electronic band structure and density of states calculations. The peak in the absorption spectrum is observed within the range from 9.30 eV to 10.10 eV. After 5.4 eV, a significant effect of pressure on absorption is seen. No significant optical anisotropy is observed. The material's absorption coefficient is very high in the ultraviolet (UV) region.

### 3.5.3 Reflectivity

In **Figure 11(d)**, the reflectivity versus energy graph is plotted for [100], [010], and [001] directions of polarization under 0 and 10 GPa. At 0 photon energy, the reflectivity lies between 0.57-0.62 for different pressures and polarization directions. With increasing energy, the value of reflectivity decreases, and at about 0.72 eV, it becomes about 0.34. After this, energy reflectivity starts increasing up to about 2.62 eV. Before 5 eV, the minimum value of reflectivity was found to be about 0.13. Therefore, this compound can be used as an anti-reflection coating. Beyond 5 eV, the effect of pressure on the reflectivity is noticed. The optical anisotropy is noticeable within the energy range of 2.50 eV to 3.13 eV. The significant anisotropy in the reflectivity makes it possible to identify the polarization states of incident light [**70**].

### 3.5.4 Refractive Index

The phase velocity of an electromagnetic wave inside a material can be determined by the real part of the refractive index $n$, and how the electromagnetic wave (EMW) attenuates inside the material can be understood from the imaginary part of the refractive index $k$. In **Figure 11(e) and (f)**, the real and imaginary parts of the refractive index are plotted. For $LiMgZr_2H_{12}$, the value of $n$ is very high near the zero energy. With increasing energy, the value of $n$ decreases up to about 0.63 eV. Then it starts increasing before about 1.1 eV energy of the incident photon, beyond 1.1 eV, the $n$ again starts decreasing. Below 9 eV, the lowest value of the real part of the refractive index $n$ is observed at 2.85 eV. No significant anisotropy is observed, and the effect of pressure is weak. For the imaginary part of the refractive index $k$, which measure the extinction, the maximum value is observed at low photon energy. Within the energy range from approximately 0.9 to 10.5 eV, several oscillatory peaks are observed. A very small effect of pressure is observed beyond 3.3 eV, and there is no noticeable anisotropy. After 20 eV, the value of $k$ is almost zero, which indicates very low radiation absorption.

### 3.5.5 Optical Conductivity

The optical conductivity parameter describes the energy-dependent conduction of free electrons under irradiation by incident photons [**70**]. In **Figure 11(g)**, we have plotted the real part of optical conductivity as a function of the energy of the incident photon for [100],[010], and [001] directions of polarization under 0 and 10 GPa for the $LiMgZr_2H_{12}$ compound. Optical conductivity starts rising from zero energy which indicates the absence of a band gap in the compound and is consistent with the energy-dependent absorption coefficient analysis. Thus, the optical conductivity also confirms the metallic nature of the $LiMgZr_2H_{12}$ compound. Anisotropy is not significant in the optical conductivity for this compound. The highest value in the optical conductivity is observed at about 9 eV for the [001] direction under zero pressure. And for 10 GPa, the highest value of optical conductivity is observed at about 9.60 eV for [001] direction. A small effect of pressure is observed after 5.30 eV.

### 3.5.6 Loss Factor

The energy losses that electrons encounter when interacting with the material as a result of the excitation of the plasma oscillation are estimated by the loss function, $L$. In **Figure 11(h)**, the energy-dependent loss function is plotted for [100],[001], and [010] directions under 0 and 10 GPa. Peak in the loss function is found at about 18.53 eV at 0 GPa and about 19.4 eV at 10 GPa. The characteristic plasmon resonance is correlated with the peak of the loss spectrum [**42**]. At energies above the plasma resonance, $LiMgZr_2H_{12}$ becomes transparent to the incident light and reflectivity, absorption coefficient, and extinction coefficient fall drastically.

# 4 Phonon Dynamics and Superconductivity

To evaluate the potential of $LiMgZr_2H_{12}$ as a conventional phonon-mediated superconductor, we calculated the key superconducting parameters at ambient pressure (0 GPa) and under a compression of 10 GPa.

**Figure 12** shows the representative phonon dispersions of $LiMgZr_2H_{12}$ at 0 GPa and 10 GPa. We can see that there is no negative phonon mode, indicating dynamical stability in each case. In the low-frequency region, the phonon spectrum is mainly dominated by the in-plane vibrational modes of Zr and H atoms. The mid-frequency and high-frequency regions are mainly dominated by the vibrational modes of H atoms only. Although there is no optical phonon gap observed at 0 GPa, there is an optical phonon gap at 10 GPa in $LiMgZr_2H_{12}$.

The superconducting properties of $LiMgZr_2H_{12}$ are evaluated using electron-phonon coupling (EPC) calculations. The superconducting transition temperature, $T_c$, is estimated using the Allen-Dynes-modified McMillan formula [**72**]:

$$T_c = \frac{\omega_{log}}{1.2} exp\left[\frac{-1.04(1+\lambda)}{\lambda - \mu^*(1+0.62\lambda)}\right] \quad (4)$$

where the $\mu^*$ is the repulsive Coulomb pseudopotential and can be set as a typical value between 0.10 and 0.16 [**73**]. In this work, $\mu^* = 0.10$ was adopted.

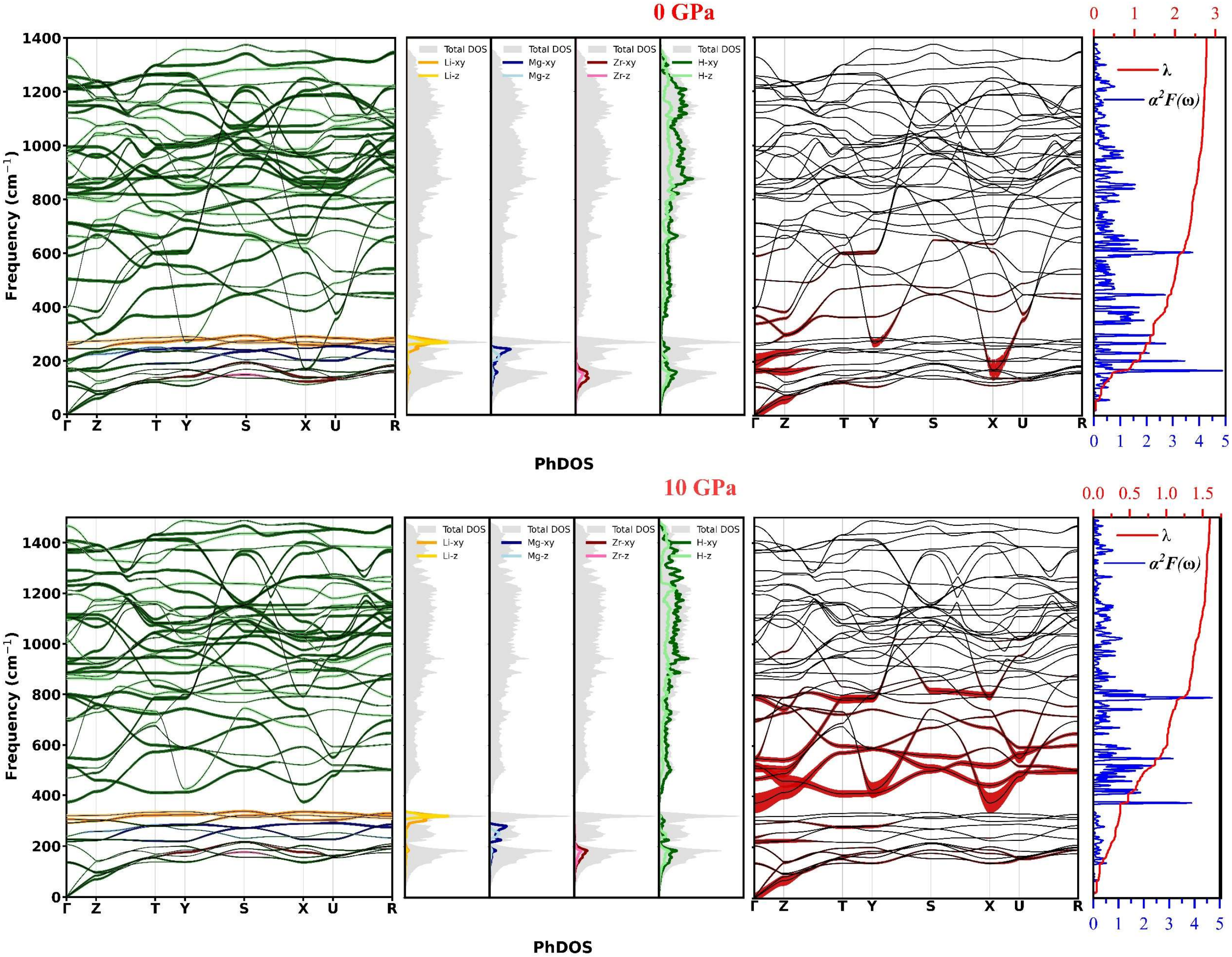

**Figure 12.** Phonon dispersions weighted by different atomic vibrational modes of $LiMgZr_2H_{12}$. The right panel displays the total phonon density of states (gray area) along with the mode-resolved contributions (colored lines). Phonon dispersions weighted by the electron-phonon coupling (EPC) strength, along with the Eliashberg spectral function, $\alpha^2F(\omega)$ and the integrated strength of EPC, $\lambda(\omega)$ for $LiMgZr_2H_{12}$ at 0 GPa and 10 GPa.

The integrated EPC constant can be evaluated as follows:

$$\lambda(\omega) = 2\int_0^\infty \frac{\alpha^2F(\omega)}{\omega}d\omega \tag{5}$$

The Eliashberg electron–phonon spectral function, $\alpha^2F(\omega)$ is written as follows [**73,74**]:

$$\alpha^2F(\omega) = \frac{1}{2\pi N(E_\mathrm{F})}\sum_{qv}\frac{\gamma_{qv}}{\hbar\omega_{qv}}\delta(\omega-\omega_{qv}) \tag{6}$$

where, $\omega_{qv}$ is the phonon frequency and $\gamma_{qv}$ is the phonon linewidth for the phonon with wave vector $q$ and branch $v$, with,

$$\gamma_{qv} = 2\pi\omega_{qv}\sum_{i,j}\int_{BZ}\frac{d^3k}{\Omega_\mathrm{BZ}}\left|g_{qv}(k,i,j)\right|^2\delta(\varepsilon_{k,i}-\varepsilon_F)\delta(\varepsilon_{k+q,j}-\varepsilon_F) \tag{7}$$

Here, $g_{qv}$ $(k, i, j)$ is the matrix element of the electron-phonon interaction, and $\varepsilon_{q,i(j)}$ is the electronic energy levels involved.

The calculated Eliasberg spectral function and the integrated strength of the EPC are plotted in **Figure 12**. As shown in the figure, the integrated strength of the EPC below the 300 $cm^{-1}$ reaches ~55 % and ~26 % of the total EPCs for the 0 GPa and 10 GPa respectively. The calculated total λ's are 2.74 and 1.57 for $LiMgZr_2H_{12}$ at 0 GPa and 10 GPa respectively. At ambient pressure, the calculated value of places $LiMgZr_2H_{12}$ firmly within the extreme strong-coupling regime. Even under 10 GPa of pressure, where the coupling strength undergoes a relative suppression to λ = 1.57, the system remains comfortably categorized as a strongly coupled superconductor. This robust, conventional strong-pairing mechanism is the primary foundation supporting the high critical temperatures observed in this material.

**Table 9.** Calculated electronic DOS at the Fermi level $N(E_F)$ (in electron states/eV. unit cell), strength of EPC (λ), $\boldsymbol{\omega_{log}}$, and superconducting transition temperature ($T_c$ in K) of $LiMgZr_2H_{12}$ at 0 GPa and 10 GPa.

| Pressure | $N(E_F)$ | λ | $\omega_{log}$ | $T_c$ |
|---|---|---|---|---|
| 0 GPa | 1.03 | 2.74 | 421.68 | 72.8 |
| 10 GPa | 0.89 | 1.57 | 651.13 | 77.3 |

The parameter $\omega_{log}$ in **Equation (4)**, is the logarithmically averaged characteristic phonon frequency, which is defined as:

$$\omega_{log} = \exp\left[\frac{2}{\lambda}\int\frac{d\omega}{\omega}\alpha^2 F(\omega)\ln\omega\right] \tag{8}$$

Taking the computed λ and $\omega_{log}$ into **Equation (4)**, one can obtain the superconducting transition temperature. The calculated $T_c$ as well as superconducting parameters of λ and $\omega_{log}$ are listed in **Table 9**. The phonon dispersions weighted by the magnitude of EPC are also drawn in **Figure 12**.

At ambient pressure (0 GPa), $LiMgZr_2H_{12}$ exhibits robust superconducting behavior with a high transition temperature $T_c$ of 72.766 K. This high $T_c$ is primarily driven by an exceptionally strong electron-phonon coupling strength of $\lambda = 2.74$, which is supported by a density of states at the Fermi level $N(E_F)$ of 1.03 states/eV/unit cell. The logarithmic average phonon frequency $\omega_{log}$ at this pressure is calculated to be 421.681 K. Upon the application of 10 GPa of pressure, the system undergoes notable electronic and vibrational changes. The applied pressure leads to a suppression of the electronic states at the Fermi level, with $N(E_F)$ decreasing to 0.89 states/eV/unit cell. This decrement has a negative effect on the electron-phonon coupling constant [**75**]. In the standard McMillan-Allen-Dynes framework, such a prominent decrease in λ would typically be expected to heavily suppress the superconducting transition temperature. However, this reduction in coupling strength is entirely counteracted by the pronounced pressure induced hardening of the phonon lattice. The logarithmic average phonon frequency $\omega_{log}$ experiences a substantial increase from 421.68 K at 0 GPa to 651.13 K at 10 GPa. Because $T_c$ is directly proportional to the prefactor $\omega_{log}$ in the Allen-Dynes equation, this massive frequency hardening overcompensates for the diminished electron-phonon coupling. Consequently, the overall superconducting transition temperature $T_c$ increases to 77.3 K at 10 GPa.

# 5 Conclusions

In this work, we demonstrate the dual functionality of the quaternary hydride $LiMgZr_2H_{12}$ as a high-$T$c superconductor and a potential hydrogen storage material. Our calculations reveal that $LiMgZr_2H_{12}$ is mechanically stable and exhibits robust superconducting behavior at ambient pressure. The predicted superconducting critical temperature is 72.8 K at 0 GPa, which increases to 77.3 K at 10 GPa. This enhancement is mainly associated with pressure-induced phonon hardening, indicating that moderate pressure can further improve its superconducting performance. Distinct van Hove singularities are observed near the Fermi level and at several high-symmetry points, which may significantly enhance superconductivity.

In addition, $LiMgZr_2H_{12}$ possesses a hydrogen storage capacity of approximately 5.36 wt%, indicating its potential relevance for energy storage applications. Qun Wei *et al.* proposed a possible synthesis route for $LiMgZr_2H_{12}$ through the reaction $LiMgZr_2H_{12} \rightarrow MgH_2 + LiH + 2ZrH_2 + 5/2\ H_2$ [**1**].

The compound exhibits useful thermomechanical characteristics for engineering applications. The optical study suggests its potential to be used as an efficient UV absorber and as anti-reflection coating. The compound exhibits ductile behavior and high machinability, making it a promising candidate for practical applications. Overall, the findings presented in this work are expected to encourage further theoretical and experimental investigations of $LiMgZr_2H_{12}$ and related hydrogen-rich metallic systems.

## Data availability

The data sets generated and/or analyzed in this study are available from the corresponding author on reasonable request.

## Declaration of interest

The authors declare that they have no known competing financial interests or personal relationships that could have appeared to influence the work reported in this paper.

## CRediT authorship contribution statement

**Jubair Hossan Abir:** Conceptualization, Software, Methodology, Formal analysis, Data curation, Visualization, Writing-original draft, review & editing; **Tauhidur Rahman:** Software, Methodology, Formal analysis, Data curation, Visualization, Writing- draft, review & editing**;** **Salauddin Muhammad Anis:** Formal analysis, Data curation, Visualization, Writing- draft, review & editing; **Saleh Hasan Naqib:** Conceptualization, Supervision, Validation, Administration, Writing- Reviewing and Editing; **Raihana Shams Islam:** Conceptualization, Supervision, Validation, Administration, Writing- Reviewing and Editing.